\documentclass[
twocolumn,
tighten,
trackchanges,
twocolappendix
]{aastex6}

\definecolor{xlinkcolor}{cmyk}{1,.5,0,0}

\hypersetup{
	pdftitle={Tidally induced bars in dwarf galaxies on different orbits around a Milky Way-like host},
	pdfauthor={Gajda et al.},
	linkcolor=xlinkcolor,     
	citecolor=xlinkcolor,     
	filecolor=xlinkcolor,  
	urlcolor=xlinkcolor,
}

\usepackage{amsmath}
\usepackage{bm}

\slugcomment{ApJ, submitted 8 March 2017, accepted 19 May 2017}

\shorttitle{Bars in dwarfs around the Milky Way}
\shortauthors{Gajda et al.}

\begin{document}

\title{Tidally induced bars in dwarf galaxies \\
on different orbits around a Milky Way-like host
}
\author{Grzegorz Gajda\altaffilmark{1,2}, Ewa L. {\L}okas\altaffilmark{1} and E. Athanassoula\altaffilmark{2}}

\altaffiltext{1}{Nicolaus Copernicus Astronomical Center, Polish Academy of Sciences, Bartycka 18, 00-716 Warsaw, Poland}
\altaffiltext{2}{Aix Marseille Univ, CNRS, LAM, Laboratoire d'Astrophysique de Marseille, Marseille, France }

\begin{abstract}
Bars in galaxies may develop through a global instability or due to an interaction with another system.
We study bar formation in disky dwarf galaxies orbiting a Milky Way-like galaxy.
We employ $N$-body simulations to study the impact of initial orbital parameters: the size of the dwarf galaxy orbit and the inclination of its disc with respect to the orbital plane.
In all cases a bar develops in the center of the dwarf during the first pericenter on its orbit around the host.
Between subsequent pericenter passages the bars are stable, but at the pericenters they are usually weakened and shortened.
The initial properties and details of the further evolution of the bars depend heavily on the orbital configuration.
We find that for the exactly prograde orientation, the strongest bar is formed for the intermediate-size orbit.
On the tighter orbit, the disc is too disturbed and stripped to form a strong bar.
On the wider orbit, the tidal interaction is too weak.
The dependence on the disc inclination is such that weaker bars form in more inclined discs.
The bars experience either a very weak buckling or none at all.
We do not observe any secular evolution, possibly because the dwarfs are perturbed at each pericenter passage.
The rotation speed of the bars can be classified as slow ($R_\mathrm{CR}/l_\mathrm{bar}\sim2-3$). We attribute this to the loss of a significant fraction of the disc's rotation during the encounter with the host galaxy.
\end{abstract}


\section{Introduction}
Bars are among the most prominent features of disc galaxies. Their share in the galaxy population depends on the
criteria employed. In the local Universe bars are hosted by at least $25\%$ of disc galaxies \citep{masters11,
cheung13}. If one includes also weak bars, the fraction may be as high as $60\%$. 
Looking into the past, the fraction of strong bars declines, down to $10\%$ at $z\approx 0.8$
\citep{sheth08}. There are some indications that the number of barred galaxies may depend on the environment, in
particular it may be higher in denser regions \citep{skibba12, mendez_abreu12}. \citet{janz12} found that bright dwarf
galaxies in the Virgo Cluster exhibit a bar fraction of $18\%$.

The subject of bar physics is very broad and here we recount only the most important facts. For more details, the
reader can refer to recent reviews of \citet{athanassoula13} and \citet{sellwood14}. One of the possible ways to create
a barred galaxy is via global instability occurring in a cold disk. Already in the first $N$-body simulations it was
found that discs of galaxies are often unstable \citep{miller70, hohl71, ostriker_peebles73, miller_smith79} and prone to the
formation of a bar-like feature in their centers.
From the analysis of the orbital structure of such objects \citet{contopoulos80} concluded that bars should be smaller than the corotation radius $R_\mathrm{CR}$.
\citet{athanassoula80}, analysing response of galaxies to forcings of different extent, also reached a similar conclusion.
Later, the simulations of \citet{athanassoula92a, athanassoula92b} confirmed this upper limit to the bar length and also added a lower limit, suggesting
that the ratio
of the corotation radius to the bar length $\mathcal{R}=R_\mathrm{CR}/l_\mathrm{bar}$ should be in a range
$1<\mathcal{R}<1.4$.
About two-thirds of galaxies with determined pattern speed are fast and almost all of them exhibit $\mathcal{R}<2$ \citep{corsini11, font17}.

Shortly after formation, bars undergo a vertical instability called buckling
\citep{combes_sanders81, combes90, raha91}. It leads to the thickening of the bar and is responsible for the development
of the boxy/peanut bulges in disc galaxies \citep[see][for a review]{athanassoula16}. \citet{athanassoula02,
athanassoula03} showed that the further evolution of the bar is governed by the transfer of the angular momentum, which
is emitted by the resonances in the bar region and absorbed by the resonances in the outer parts of the disc and particularly in the dark matter halo. As a result, the bar slows down and is able to grow. 
The formation of bars was also studied in hydrodynamical cosmological simulations \citep{scannapieco_athanassoula12, algorry16}.

Bars may also form in response to an interaction with a perturber, which was first studied by \citet{noguchi87} and \citet{gerin90}.
Such bars seem to be similar to the ones formed in isolation, however, there are two important differences. If the
galaxy is initially sufficiently stable against spontaneous bar formation, the pattern speed of a tidally induced bar is not
decreasing in a secular fashion, but remains constant \citep{salo91, miwa_noguchi98, martinez_valpuesta17}. Moreover,
such bars are usually slow, having $\mathcal{R} \sim 2-3$ \citep{miwa_noguchi98, berentzen04, aguerri_gonzalez_garcia09, lokas14, lokas16}.
\citet{miwa_noguchi98} argued that such values of $\mathcal{R}$ are caused by the transfer of the angular momentum to the perturber.

Encounters between galaxies have been known for a long time to change the appearance and structure of galaxies
\citep[e.g.][]{toomre_toomre72}. Tidal force influences the evolution of galaxies in groups \citep{villalobos12} and
clusters \citep{mastropietro05, lokas16, semczuk17}. In the tidal stirring scenario, disky dwarf galaxies are
transformed into dwarf spheroidals, both in the case of dwarfs around normal-size galaxies \citep{mayer01, kazantzidis11}
and in clusters \citep{aguerri_gonzalez_garcia09}. During the first encounter with the host galaxy, a bar develops in
the disc of the dwarf \citep{klimentowski09}. At subsequent pericenter passages the disc thickens and later on it
is transformed into a spheroid.

If the tidal stirring scenario is valid, we may expect to find some dwarfs with hints of bar presence in the Local
Group. There is evidence that the elongated shape of the Sagittarius, Ursa Minor and Carina dwarfs \citep{lokas10,
lokas12, fabrizio16} may be the result of the bar phase they underwent during their tidal evolution. In addition,
also the ultra-faint dwarfs Hercules and Ursa Major II exhibit bar-like shapes \citep{coleman07, munoz10}. Of course,
the tidally induced bars may form not only in the Local Group, but also in other dwarf galaxies interacting with their
hosts. One of the possible examples is a small galaxy in the Arp 83 system (also known as NGC 3799), which seems to
interact with a larger host NGC 3800.

Recently, \citet{lokas14} studied in detail the evolution of a tidally induced bar in a dwarf orbiting around a
Milky Way-like galaxy. Here, we extend their results to different orbital configurations. The main issue we focus on is
the impact of two parameters on the bar formation. We study the outcome of varying the orbit size (i.e. peri- and
apocentric distances) and the inclination of the dwarf's disc with respect to the orbit. We also draw some more general
conclusions regarding tidally induced bars.
The paper is organized as follows. In Section \ref{sec_simulations} we introduce our simulations. In Section
\ref{sec_results} we describe the results. Next, in Section \ref{sec_discussion} we discuss the results and we
summarize our work in Section \ref{sec_summary}.

\section{Simulations}
\label{sec_simulations}

\begin{figure}
\includegraphics[width=\columnwidth]{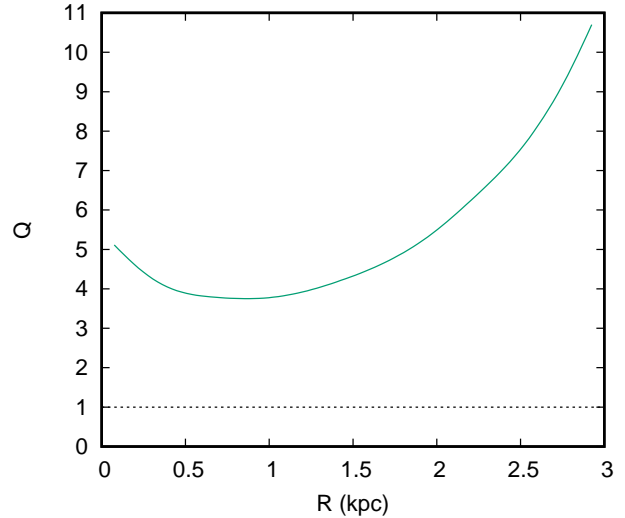}
\caption{
Initial profile of the Toomre parameter for the dwarf galaxy. The disc is stable if $Q>1$.}
\label{fig_q_param}
\end{figure}

\begin{table}
\centering
\caption{Basic properties of the simulations.}
\label{tab_simulations}
\begin{tabular}{ccccc}
\hline\hline
Run & Apocenter & Pericenter & Inclination & Line \\
 & (kpc) & (kpc) & (deg) & color \\
\hline
M0 & 120 & 24 & 0  & Black    \\
S0 & 100 & 20 & 0  & Green  \\
L0 & 250 & 50 & 0  & Blue   \\
M45 & 120 & 24 & 45 & Violet \\
M90 & 120 & 24 & 90 & Red \\
\hline
\end{tabular}
\end{table}

Our simulation setup resembles the one adopted by \citet{lokas14}. We constructed $N$-body models of the dwarf and
the host galaxy.
Both consist of a spherical NFW \citep{nfw95} dark matter halo, which is exponentially truncated at the virial radius, and an exponential stellar disc.
The models were generated using procedures described in \citet{widrow_dubinski05} and \citet{widrow08}. Each component
of each galaxy was made of $10^6$ particles. According to \citet{dubinski09} such a number of particles is sufficient to faithfully reproduce the evolution of a barred galaxy.

To model the dwarf, we used a dark matter halo of the $10^9$ M$_\sun$ virial mass and a concentration of $20$.
The exponential stellar disc has a mass of $2\times 10^7$ M$_\sun$, a radial scale-length of $0.41$ kpc and a vertical
scale-length of $0.082$ kpc. The host galaxy was designed to resemble the Milky Way. Its dark
matter halo virial mass was set to $7.7\times 10^{11}$ M$_\sun$ and its concentration to $27$. The exponential disc has a mass of
$3.4\times 10^{10}$ M$_\sun$, the radial scale-length is $2.82$ kpc and the vertical scale-length is $0.44$ kpc.

We constructed the dwarf galaxy model ensuring that it is stable against spontaneous bar formation. \citeauthor{toomre64}'s \citeyearpar{toomre64} stability criterion reads
\begin{equation}
Q=\frac{\sigma_R \kappa}{3.36 G \Sigma} > 1,
\end{equation}
where $\sigma_R$ is the radial velocity dispersion, $\kappa$ is the epicyclic frequency, $\Sigma$ is the surface density of the disk and $G$ is the gravitational constant. We plotted the initial profile of $Q(R)$ for the dwarf galaxy in Figure~\ref{fig_q_param}.
The minimum value is equal to $3.75$, hence no bar should grow in its disk for times comparable to the evolution times considered here.
We verified this by running an additional $10$ Gyr simulation of the dwarf galaxy in isolation.
The model of the host galaxy has a similar property, having a minimum value of $Q\approx 2$.
However, the bar in the host would probably have no impact on the gravitational potential felt by the dwarf on its orbit.

We intended to gauge the impact of two factors on the bar formation: the size of the orbit of the dwarf galaxy and the inclination of its stellar disc with respect to the orbit.
We chose a fiducial simulation of a pericenter $24$ kpc, an apocenter $120$ kpc and fully in-plane prograde orientation of the disc.
It means that in the beginning the disc lies in the same plane as the orbit of the dwarf and rotates in the same direction as it orbits the host galaxy.
Such an orientation maximizes the impact of the tidal interaction  \citep{kazantzidis11, lokas15}. We will refer to this simulation as run
M0 (where the letter `M' stands for medium-size orbit and the number `0' indicates the inclination angle in degrees).

In the next two runs we changed the size of the orbit, keeping the ratio of the apocenter to the pericenter distance
equal to the typical values of $5$, as well as the in-plane prograde orientation. In the run S0 (`S' for small) the orbit was
tighter than M0, with the pericenter of $20$~kpc and the apocenter of $100$ kpc. The L0 (`L'~for large) orbit was
wider, with the pericenter at $50$ kpc and the apocenter at $250$ kpc.

In two other runs we varied the inclination angle of the dwarf's disc, while keeping the same, medium, orbit size.
Compared to run M0, we rotated the disc around the line connecting the initial position of the dwarf and the host
center in anticlockwise direction, as viewed from the host. In run M45 we rotated the disc by $45\degr$, while in
run M90 the disc was rotated by $90\degr$, so it was perpendicular to the plane of the orbit.
The simulation runs are summarized in Table \ref{tab_simulations}. We listed the designations of the runs, the apo- and
pericenter distances, as~well as the initial disc inclinations. In the last column we give colors, which we will use
for a given simulation throughout the paper.

To give the simulations a gentle start, we always initially place the dwarfs at the apocenters of their orbits.
The initial orbital velocity is set so as to reach the desired distance at the pericenter.
Throughout the simulation runs, the orbits of the dwarfs do not decay significantly.
The pericenter distances remain almost constant, while the apocenters decrease by less than $10\%$.

To run the simulations we used the publicly available code \textsc{gadget2} \citep{springel05}. The evolution of the
system was followed for $10$ Gyr and outputs were saved every $0.05$ Gyr. For the dwarf we adopted the following
softening lengths: $0.02$ kpc for the stellar particles and $0.06$ kpc for the dark matter particles. In the case of the
host galaxy, we used $0.05$ kpc for the stellar component and $2$ kpc for the dark matter halo.

\section{Results}
\label{sec_results}

\subsection{The shape of the stellar component}
\label{sec_shape}

\begin{figure*}
\includegraphics[width=0.5\textwidth]{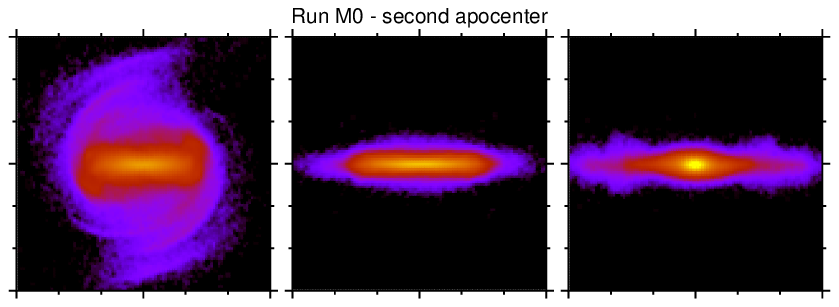}
\includegraphics[width=0.5\textwidth]{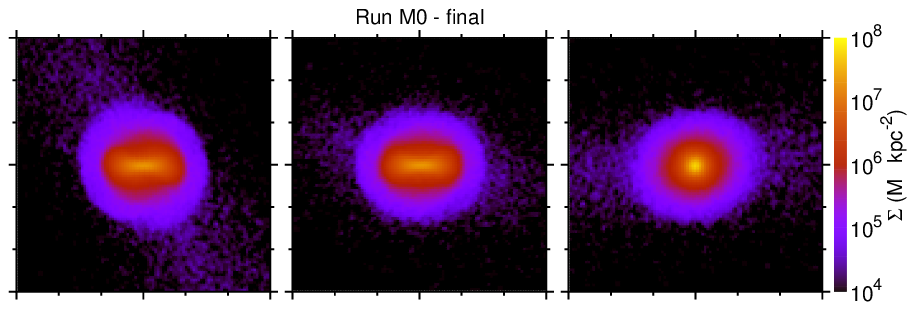} \\
\includegraphics[width=0.5\textwidth]{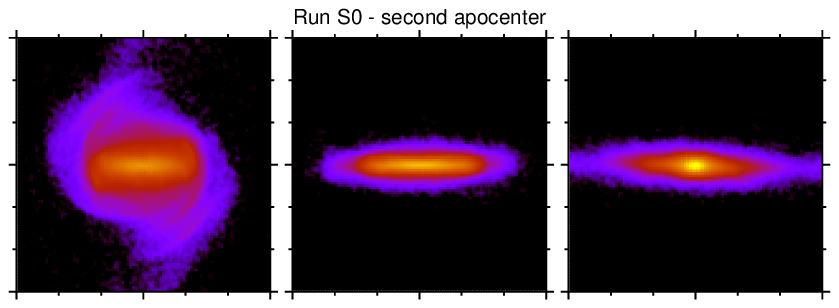}
\includegraphics[width=0.5\textwidth]{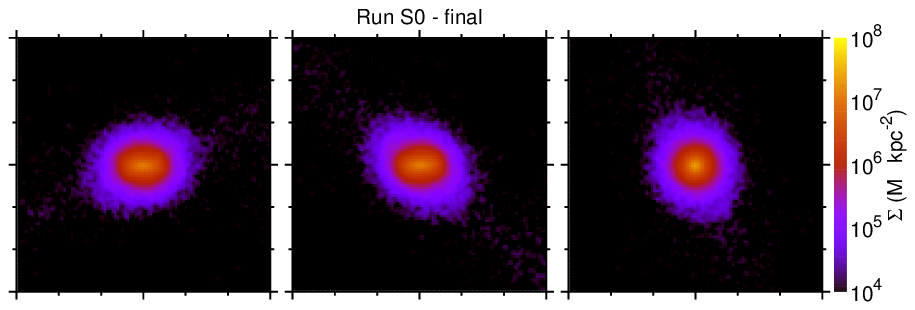}\\
\includegraphics[width=0.5\textwidth]{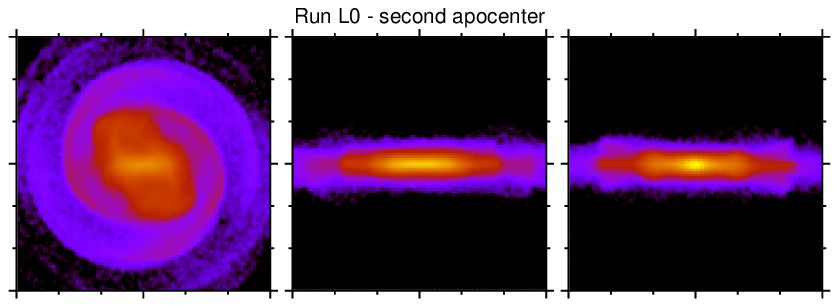}
\includegraphics[width=0.5\textwidth]{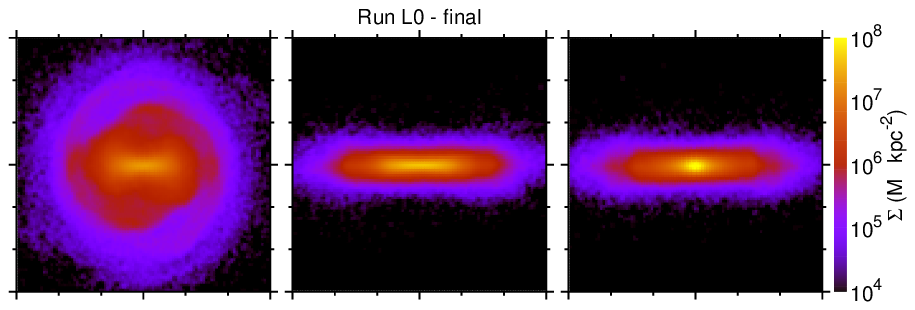}\\
\includegraphics[width=0.5\textwidth]{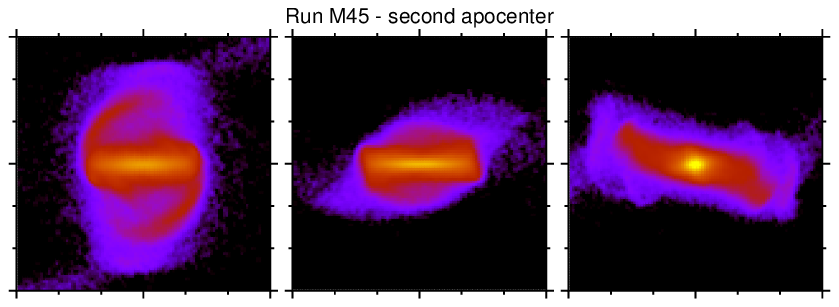}
\includegraphics[width=0.5\textwidth]{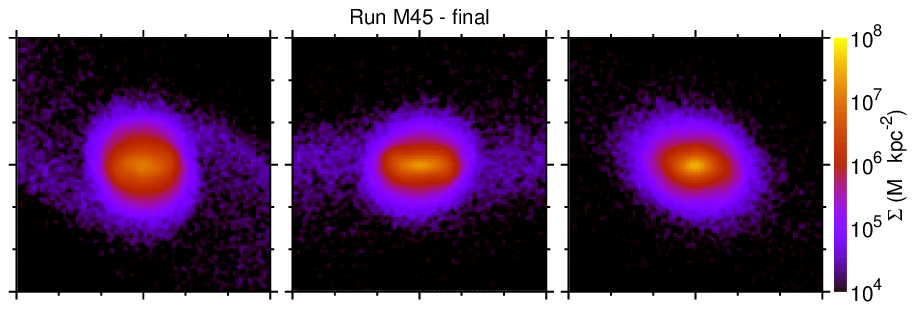}\\
\includegraphics[width=0.5\textwidth]{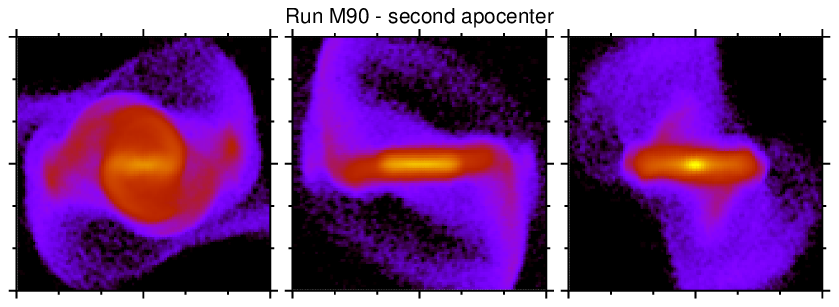}
\includegraphics[width=0.5\textwidth]{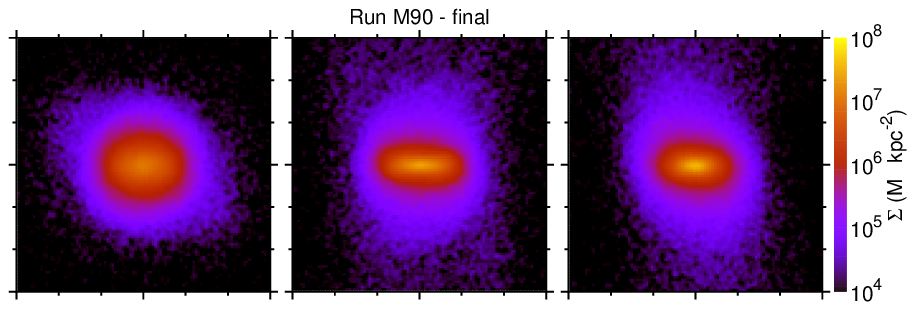}
\caption{Surface density maps of the stellar component of the dwarf galaxies.
Each row corresponds to one of the simulation runs.
The set of first three columns presents the $(x,y)$, $(x,z)$ and $(y,z)$ views of the dwarfs at their second apocenter (i.e. after the first pericenter passage), respectively: the face-on, edge-on and end-on view.
The second set of three columns shows the same three views, but after $10$ Gyr of evolution. The side of each panel is $6$ kpc. Small ticks are $1$ kpc apart. }
\label{fig_density_maps}
\end{figure*}

\begin{figure*}
\includegraphics[width=\textwidth]{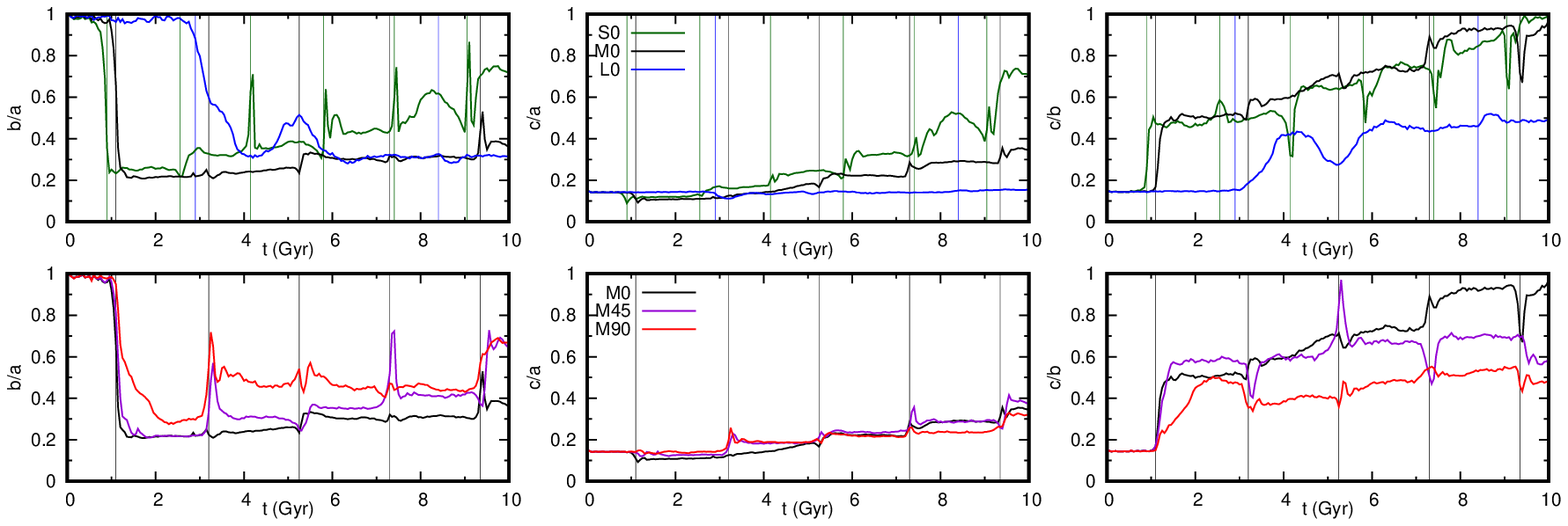}
\caption{Axis ratios of the dwarf galaxy as a function of time. From left to right: minor to major, intermediate to major and minor to
intermediate. Top row: impact of the orbit size. Bottom Row: impact of the inclination. Vertical lines indicate
pericenter passages and are shown in colors corresponding to a given simulation.}
\label{fig_shape}
\end{figure*}

In the following we will be using a reference frame aligned with the bar. The principal axes of the bar were determined
through diagonalization of the inertia tensor \citep{zemp11, gajda15}, calculated for the particles inside an ellipsoid
of semi-major axis length equal to $1$ kpc. We rotate the reference frame in such a way that the $x$-axis is aligned
with the major axis of the bar, the $y$-axis with the intermediate and the \mbox{$z$-axis} with the minor one. In this task, as
well as for computing other quantities, we used the scale of $1$ kpc, which is slightly smaller than our typical bar
lengths. In principle, we could have used the bar length itself for this purpose, however we decided to refrain from
this because it would make it difficult to disentangle the intrinsic changes of a given quantity and the variation due
to the change of the bar length. Moreover, the length of the bar itself is not an easily defined quantity
\citep{athanassoula_misiriotis02}.

In Figure \ref{fig_density_maps} we show surface density maps of the dwarfs after their first pericenter passage
(strictly speaking, at~their second apocenters) and at the end of the simulations (after $10$ Gyr).
At the latter time, all of them are close to their next apocenter, so their shapes are not very disturbed. In all the dwarfs, bars
are clearly visible after the first pericenter passage, but their lengths and shapes are different. There are no
peanut shapes visible in the edge-on views, hence it seems that the dwarfs did not undergo recognizable buckling. The
end products of the evolution are also diverse but in all the dwarfs the central stellar distributions remain elongated.

First, let us examine the influence of the orbit size. In the course of the simulation, the fiducial dwarf M0 became
almost spherical and only its center remained elongated.
Early in its life, the bar formed in dwarf~S0 (on the tighter orbit) is
slightly less extended than the bar in M0. At the end, the dwarf is smaller and rounder due to enhanced tidal stripping.
The bar in run L0 (wider orbit) is also shorter than M0. An elongated shape visible in the face-on view at the second apocenter, inclined to the bar in
the center, originates from winding up of spiral arms and will be discussed later on. This run is the only one in which
the dwarf remained clearly disky until the end, with an easily discernible bar. It is the result of the fact that the L0
dwarf experienced only two pericenter passages.

In the case of the inclined dwarfs (M45 and M90) there is plenty of material out of the disc plane after the first
pericenter. It was torn off the disc by the tidal force because both discs are inclined with respect to the plane of
the orbit. The bars get shorter with growing inclination, however the final shapes of the dwarfs M45 and M90 are qualitatively similar to the M0 case.

The total amount of stripped material strongly correlates with the size of the orbit.
Thus, the dwarf on the S0 orbit lost the largest amount of matter, as expected, while the one on L0 the least.
The dwarfs on the same orbit (M0, M45, M90) lost a similar fraction of mass, but the trend between them is such that the fraction drops with the inclination \citep[see also][]{lokas15}.

Figure \ref{fig_shape} depicts the evolution of the dwarf stellar component shape. We quantify it in terms of the axis
ratios: intermediate to major ($b/a$), minor to major ($c/a$) and minor to intermediate ($c/b$). In the calculations of
the axis ratios we used stellar particles inside ellipsoids of semi-major axis equal to $1$ kpc, hence
approximately $a=1$~kpc. This length is of the order of the bar sizes, so here we actually measure the \emph{bar}
shapes. All the dwarfs were initially axially symmetric discs ($b/a=1$), but their shapes changed abruptly at the
first pericenter passage.
This hints that the bar forms in a short time at the first pericenter, as we will show in detail in Sec. \ref{sec_bars}.
Later on, the parameters are usually constant between pericenters.
At the pericenters, the measured shape may fluctuate significantly, 
as can be seen especially in the case of run S0, in which there are large spikes in $b/a$.
This is related to the fact that at pericenter the bars are not necessarily oriented along the line from the dwarf to the host, along which the tidal force acts.
Thus, during the encounter with the host galaxy the tides may temporarily make the central parts of the dwarf rounder.
The size of this effect depends both on the orientation of the bar and the pericenter distance.

\begin{figure}
\includegraphics[width=\columnwidth]{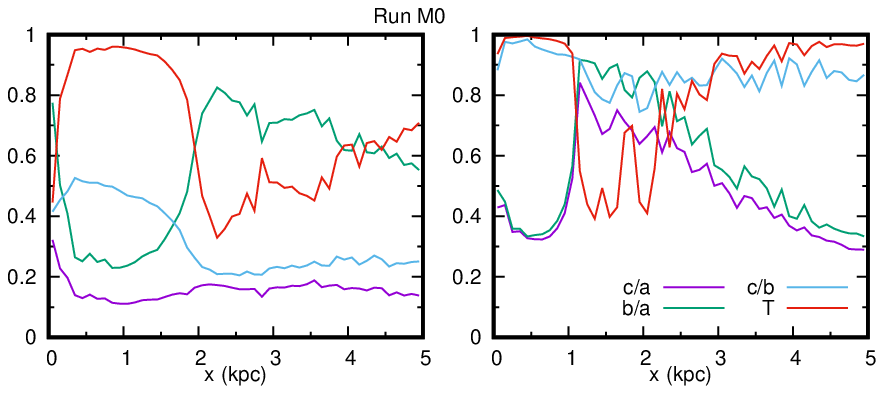} \\
\includegraphics[width=\columnwidth]{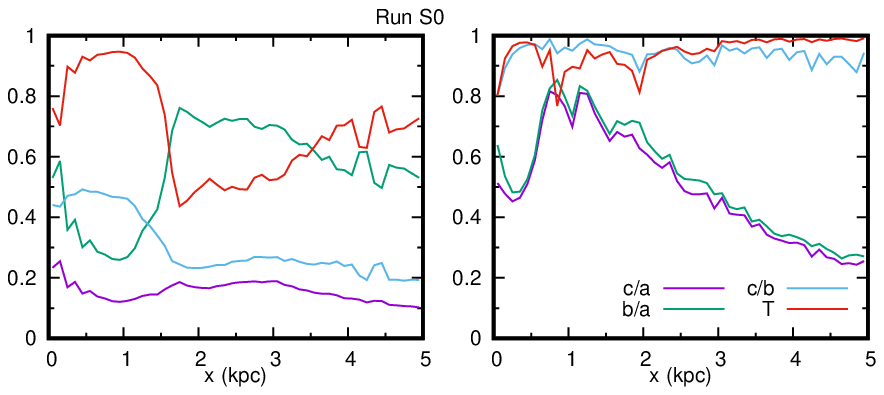} \\
\includegraphics[width=\columnwidth]{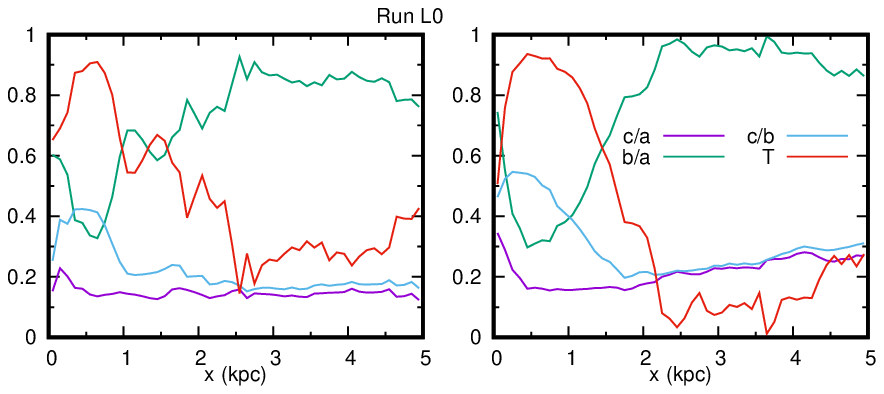} \\
\includegraphics[width=\columnwidth]{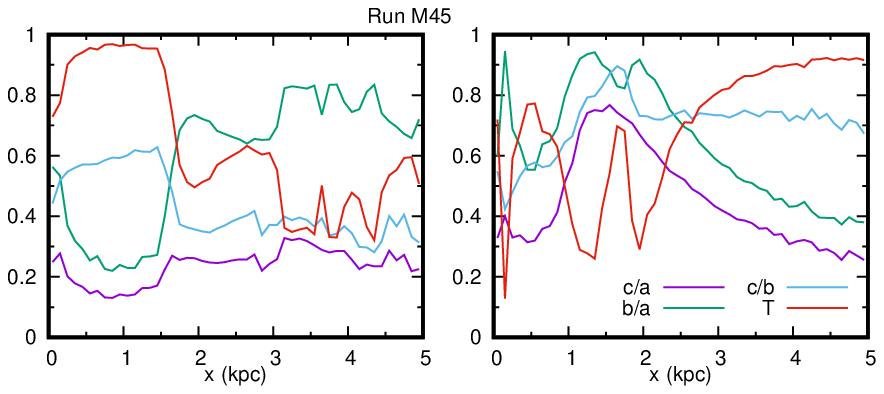} \\
\includegraphics[width=\columnwidth]{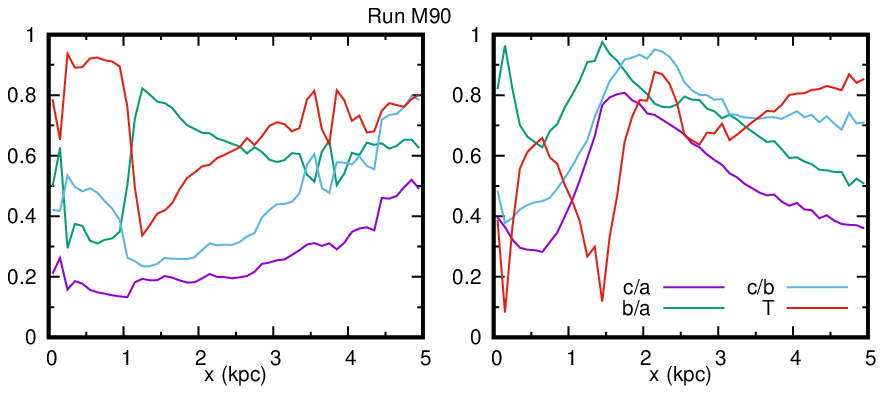}
\caption{Shapes of the stellar component at the second apocenter (left column) and at the end of the simulations
(right column) as a function of distance from the center of the dwarf. Each row corresponds to a different run. The
violet lines show the minor to major axis ratio ($c/a$), the blue ones minor to intermediate ($c/b$), and the green
ones intermediate to major ($b/a$). The triaxiality parameter, $T$, is plotted with the red line.
}
\label{fig_shape_lin}
\end{figure}

\begin{figure}
\includegraphics[width=\columnwidth]{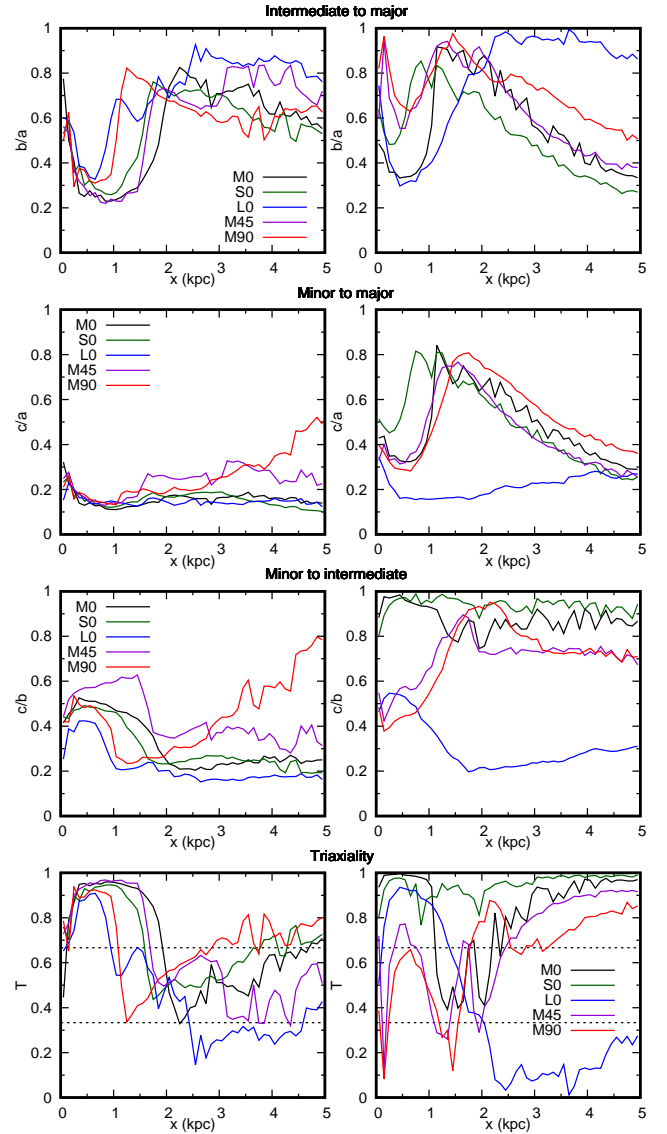} \\
\caption{Shapes of the stellar component at the second apocenter (left column) and at the end of the simulations
(right column) as a function of distance from the center of the dwarf.
The data are the same as for Figure \ref{fig_shape_lin}, but the layout is such as to allow comparisons between runs. From top to bottom,
the rows depict intermediate-to-major, minor-to-major and minor-to-intermediate axis ratios and finally triaxiality.
In the last row, dashed lines indicate ranges of prolateness ($T>2/3$), triaxiality ($2/3>T>1/3$) and oblateness ($T<1/3$).
Colours of the lines correspond to the different simulation runs.
}
\label{fig_shape_lin_new}
\end{figure}

After each pericenter the bar usually becomes less elongated, but the strength of this effect depends on the size of
the orbit. Also the thickness increases during the encounters with the host. Interestingly, in the M0 run, during the
period between $3.5$ and $5.5$ Gyr, $c/a$ grows steadily, suggesting a possible buckling episode. Upon detailed
inspection of the density maps, this turned out to be true, however it was barely noticeable. We note the bars
initially have $c/b \sim 0.5$, indicating they are not axisymmetric, prolate ellipsoids. However, the dwarfs S0 and M0
reach such a state at the end of the runs.

The impact of the inclination on the elongation is such that for a more inclined disk the bar is initially less
elongated and remains so. On the other hand, the change in thickness (measured by $c/a$) is similar for all the dwarfs
M0, M45, and M90.

We used the methods  mentioned earlier \citep{zemp11, gajda15} to measure how the shape of the stellar component varies
with the distance from the center of the dwarf. According to \citet{zemp11}, the axis ratios obtained by this method
correspond to ellipsoidal shells of constant density. In addition to the axis ratios, to quantify the shape we also use
a triaxiality parameter $T$, defined as
\begin{equation}
	T=\frac{1-(b/a)^2}{1-(c/a)^2}.
\end{equation}
An object with $T<1/3$ can be considered oblate (disky) and if $T>2/3$ it is prolate (cigar-like). For the
intermediate values of $T$ the shape is triaxial. In Figures \ref{fig_shape_lin} and  \ref{fig_shape_lin_new} we present the axis ratios and $T$ as a
function of distance along the major axis. As in the case of the surface density maps (Figure~\ref{fig_density_maps}),
we show the results when the dwarfs were at the second apocenter and at the end of the simulations.

After the first pericenter passage the profiles of the axis ratios are qualitatively similar for all dwarfs. The discs
remained thin ($c/a \approx 0.15$), with an exception of the outer parts of the inclined runs (M45 and especially M90),
which is caused by the material pulled out of the disc plane by the tides. The profiles of the intermediate-to-major
ratios underwent considerable changes. In the central parts $b/a$ dropped to $\approx 0.3$, signifying the presence of a
bar. In the outskirts it grows again, but only to $\approx 0.8$, hence the discs are not fully round.

The shape evolution is well described by the triaxiality, $T$. In the bar region $T\approx 0.9$, as expected for an~elongated bar. Further out, it drops abruptly, which corresponds to the end of the bar. However, it barely drops to
$\sim 1/3$, hence the shape is not fully disky, but rather triaxial. The prominent exception is run L0, whose outer
parts remained disky because of the large pericenter distance. The very central parts ($x<0.25$ kpc) appear to be
rounder than the bars. We caution that it might be the effect of insufficient resolution. An ellipsoid of major axis
$a=0.2$ kpc and $b/a=0.3$ would have $b=0.06$ kpc, which is only thrice the softening length of the stars
and of the order of the softening of dark matter particles.

The most elongated and the longest bar developed in dwarf M0. Both smaller and larger pericenter distances lead to a
shorter bar. Higher inclination of the disc also leads to bar suppression.

The shapes of the dwarf galaxies at the end of the runs are different and depend strongly on the initial conditions. In
runs M0 and S0 the stellar component in the center became axisymmetric ($c/b \approx 1$), while remaining prolate.
The elongated part is smaller and thicker than in the beginning, while its outskirts developed into a spherical
envelope. At the end, S0 is less elongated and more axisymmetric than M0 and we can consider it to be more evolved
due to a larger number of closer pericenter passages. As we have seen before, run L0 is different, with bar becoming
more prominent and longer. In addition, its outer parts retained the shape of a thin disc.

In the case of the inclined discs we can see a different progression at the end of the simulations. While the thickness
profiles ($c/a$) of runs M45 and M90 resemble the one of run M0, the ratio $b/a$ is different. The shape in the center
is considerably rounder, as indicated by a larger $b/a$ value. As a result, $T$ has a smaller maximum. Actually,
according to our criteria, the bar in run M90 does not have a prolate shape ($T_\textrm{max}<2/3$). However, the region
where the distribution is elongated is distinguishable from the more spherical envelope.
In most cases, $T$ grows at large distances. This indicates the transition from the main body of the dwarf to its
tidal tails, which are obviously elongated.

\begin{figure}
\includegraphics[width=\columnwidth]{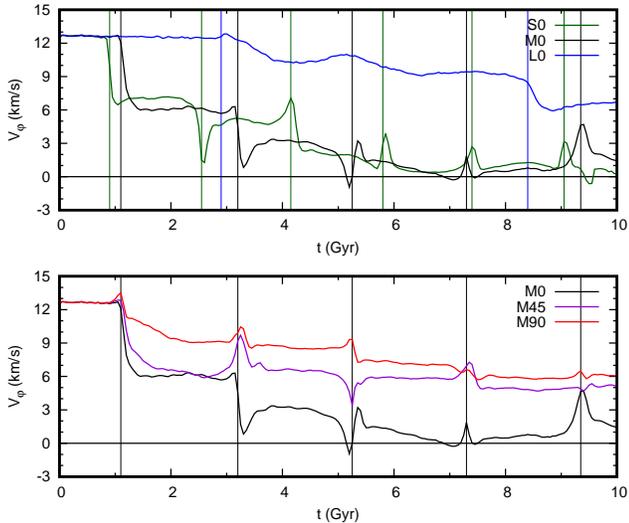}
\caption{Evolution with time of the rotation velocity of the stellar component. Vertical lines indicate pericenter passages.}
\label{fig_rotation}
\end{figure}

\newpage
\subsection{Stellar kinematics}

To study the kinematics of the stellar component, we will use a spherical coordinate system including the distance from
the center $r$, the polar angle $\theta$ and the position angle $\varphi$. In Figure \ref{fig_rotation} we show the
mean rotation velocity $V_{\varphi}$ calculated including all particles inside a~$1$~kpc sphere.
All dwarfs lost rotation due to tidal interaction with the host galaxy.
The strong decrease at the first pericenter is an effect of the bar formation inside the discs.
This decrease is relatively smaller in the cases of L0 and M90 because their newly developed bars are shorter and weaker than in the other runs (see~Sec.~\ref{sec_bars}).
In the course of the whole $10$ Gyr of evolution,
the most severe decrease took place for the dwarfs with small pericenters and in-plane prograde discs (S0 and M0).
Conversely, a large pericenter or an inclined disc allowed a~dwarf to retain rotation to some degree.

\begin{figure}
\includegraphics[width=\columnwidth]{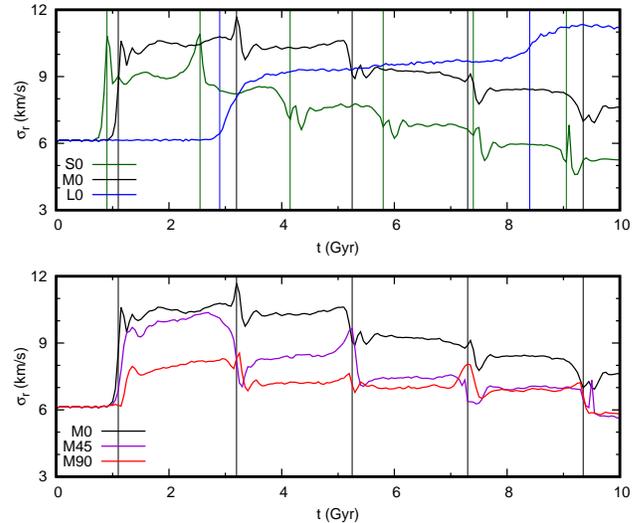}
\caption{The radial velocity dispersion of the stars, $\sigma_r$, measured inside $1$ kpc, as a function of time. Vertical lines indicate
pericenter passages.}
\label{fig_dispersion}
\end{figure}

In Figure \ref{fig_dispersion} we show the evolution of the radial velocity dispersion $\sigma_r$ measured inside $1$ kpc sphere. At~the first pericenter, it grows in all cases, indicating the emergence of more radial orbits, typical for bars.
In some cases there are spikes of $\sigma_r$ at the time of the pericenter passages, caused by strong disturbances of the dwarfs, especially in the run S0.
The average value of $\sigma_r$ after the first pericenter passage was largest
in run M0, in which the bar is also the strongest.
The later average decrease of this parameter is caused by the mass loss at subsequent pericenters passages.
An exception to this trend was dwarf L0, for which $\sigma_r$ grew at the second pericenter,
possibly because the stripping was rather weak, allowing the bar to grow. The velocity dispersion is fairly constant
between pericenters, indicating that the bar is stable at these periods of time. We checked also the dispersion of
transverse velocities $\sigma_\theta$ and $\sigma_\varphi$. The evolution of $\sigma_\varphi$ is qualitatively similar to
the evolution of $\sigma_r$, only the initial growth is smaller.
The $\sigma_\theta$ velocity dispersion is initially small and it grows later on. In the case of S0 both are almost identical in the end.

\subsection{Properties of the bars}
\label{sec_bars}

\begin{figure}
\includegraphics[width=\columnwidth]{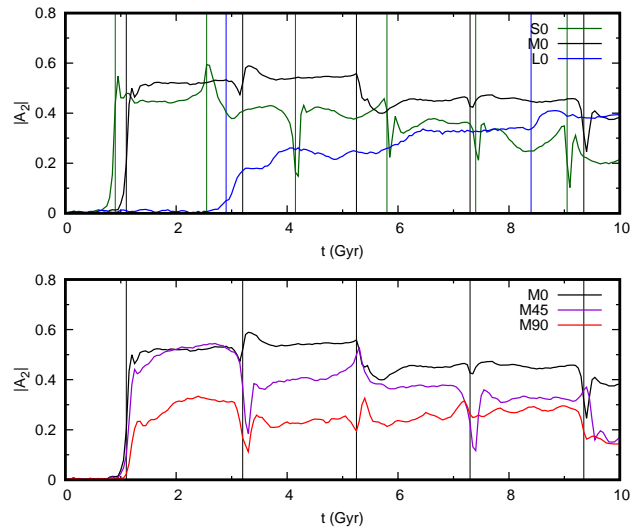}
\caption{The bar mode $|A_2|$ measured inside $1$ kpc as a~function of time. Vertical lines indicate
pericenter passages.}
\label{fig_a2_tot}
\end{figure}

Now we are turning to specific diagnostics of the bar itself.
To quantify the bar strength we are going to use the so-called \emph{bar mode}
\begin{equation}
	A_2=\frac{1}{N}\sum\limits_{j=1}^{N} \exp(i\, 2 \varphi_j),
\end{equation}
where the summation runs over all stellar particles in the region of interest, $i$ is the imaginary unit and
$\varphi_j$ is the position angle of the $j$-th particle. The bar strength is given by $|A_2|$, whereas its position
angle by $\arg(A_2)/2$. We calculated $A_2$ for all stars inside a cylinder of $1$ kpc radius, centered on the
dwarf galaxy. The results are  presented in Figure \ref{fig_a2_tot}. The bar mode
grows abruptly at the first pericenter passage when all the bars develop. Subsequent pericenters may strengthen the bar
a little, however in most cases they become weaker after each encounter with the host galaxy.

Regarding the impact of the size of the orbit, the strongest bar is formed for the default orbit M0. The smaller
pericenter distance of S0 does not lead to a~stronger bar. This may be due to a shorter time of the interaction or a
stronger disturbance of the disc. Later on, the S0 bar is weakened at each pericenter. The bar in the dwarf L0, on much
wider orbit, is born much weaker, however it gets stronger during the second pericenter passage.
The mildly inclined M45 hosts initially a bar as strong as M0, but afterwards it is weakened more. The bar in M90 dwarf
is not as strong as in M45, but until the fifth pericenter passage its $|A_2|$ is constant or even slightly growing.

\begin{figure}
\includegraphics[width=0.495\columnwidth]{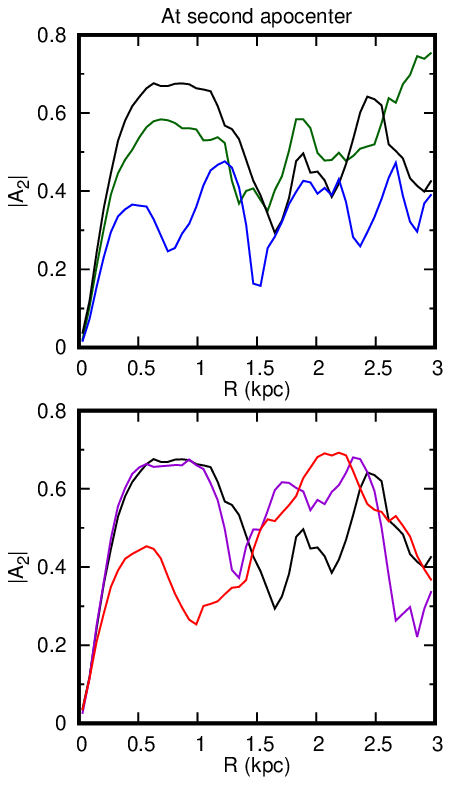}
\includegraphics[width=0.495\columnwidth]{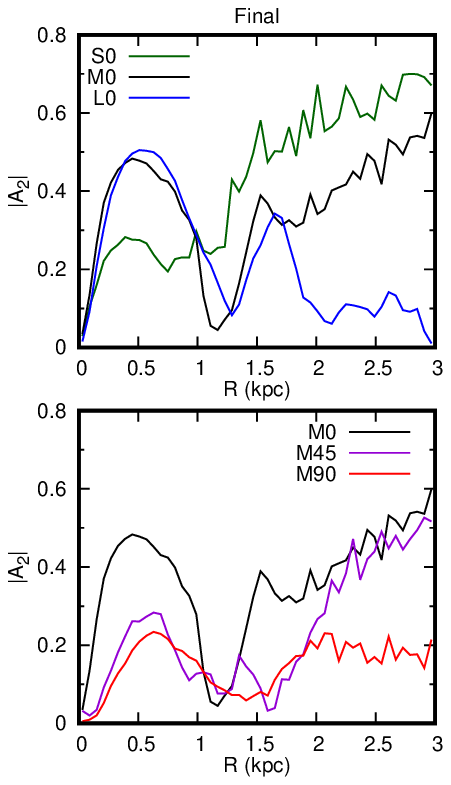}
\caption{The bar mode $|A_2|$ profiles at the second apocenter (left column) and at the end of the simulations (right
column).}
\label{fig_a2_profiles}
\end{figure}

In Figure \ref{fig_a2_profiles} we show the radial profiles of $|A_2|$ after the first pericenter passage and at the
end of the simulation runs. Obviously, the innermost peaks correspond to the bars themselves. However, the overall
profile shapes are significantly more complicated than in the case of the bars formed in isolation, for which $|A_2|$
falls down and remains low in the outskirts \citep[e.g.][]{athanassoula_misiriotis02}.
Here we have structures of larger amplitudes in the profiles.
The peaks occurring away from the centers are related to rings, spiral arms or shells of matter travelling outward.
The noisy increase of $|A_2|$ at the outskirts of the dwarfs, especially at the final stages, is
caused by the low-density tidal tails formed by the stripped material.

\begin{figure}
\includegraphics[width=\columnwidth]{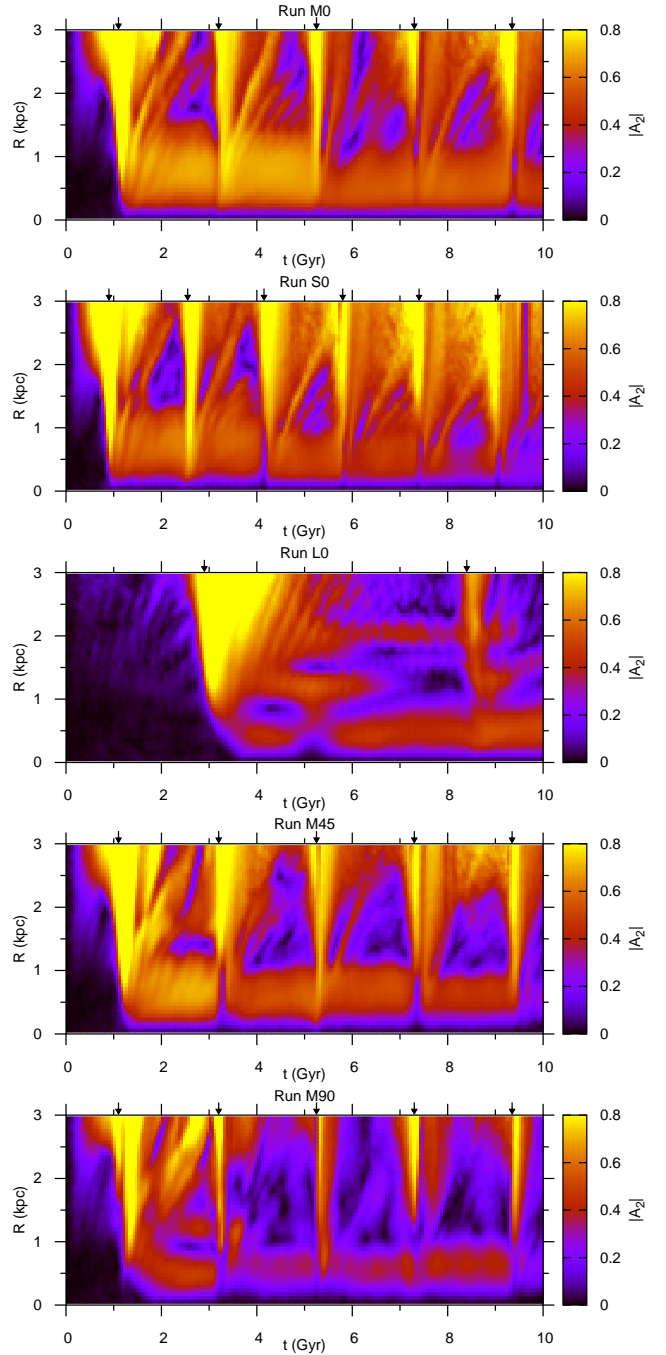}
\caption{The bar mode $|A_2|$ versus time and distance from the dwarf's center. Small arrows above the plots indicate the pericenter passages. Note that at those times the outer regions are saturated (i.e. $|A_2|>0.8$).}
\label{fig_a2_map}
\end{figure}

In Figure \ref{fig_a2_map} we illustrate how the radial profiles of $|A_2|$ evolve with time. One can immediately notice
the pericenter passages at which the outer parts of the dwarfs are strongly stretched. In all cases bars form
after the first pericenter, as indicated by the emergence of $|A_2|$ maxima at the scales of $0.5$--$1$ kpc.
The profiles of $|A_2(r)|$ does not evolve significantly in between pericenters, as can be also inferred from the global measurement in Figure \ref{fig_a2_tot}.
One can also notice narrow $|A_2|$ maxima traveling
outward. We inspected the relevant density maps and concluded that these are shells of matter or
density waves, which were pulled out during the close encounter with the host.

Out of the three in-plane prograde runs with differing orbit sizes, the longest and strongest bar is formed in run M0 with the
intermediate pericenter distance. In runs M0 and S0 the bar is getting weaker and shorter at each pericenter. The dwarf
S0 is severely stripped, as indicated by yellowish color beyond $\approx 1.5$ kpc at $8.5$~Gyr. In that region, only
some stellar particles remain, forming elongated tidal tails. On the other hand, in L0 the bar actually gets stronger
at the second pericenter passage. Examining the dependence of bar properties on the orientation of the disc, we note
a clear progression that the more inclined the disc, the weaker the bar is.

\begin{figure}
\includegraphics[width=\columnwidth]{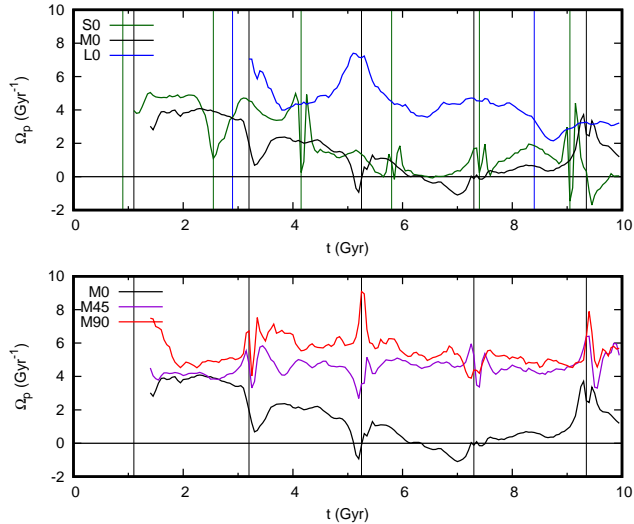}
\caption{Pattern speed $\Omega_\mathrm{p}$ of the bar as a function of time. The curves start when the bars are fully formed.}
\label{fig_omega_p}
\end{figure}

We would like to calculate the bar pattern speed $\Omega_\mathrm{p}$. Unfortunately, the discs of the dwarfs are precessing, especially in run M45, 
so that we cannot simply compare position angles of the bar at two outputs.
There are at least two possible approaches to solve this issue.
The first one involves construction of appropriate matrix operator representing transformation of the galaxy principal axes between two outputs.
The bar rotation angle can be obtained from the components of this matrix.
In the second approach one can associate the initial and final orientations of the vector normal to the disc plane and the bar with four points on a unit sphere.
The angle we are seeking can be found by means of spherical trigonometry.
We decided to take the former approach and the details of the construction are described in Appendix \ref{appendix}.
We checked that the results obtained with the second method are the same.
In fact, in our simulations the precession rates are small.
Thus, our precise measurements are not very different from results which can be obtained with some approximate method, such as measuring the 3D angle between the initial and final orientation of the bar.

We cannot use in our method the principal axes computed in Sec. \ref{sec_shape}, because some of the bars are initially shorter than $1$ kpc.
Instead, we use the $z$-axes from Sec.~\ref{sec_shape} (which correspond to the direction normal to the disc plane) and compute the bar orientation from the phase of $A_2$, calculated using particles inside cylinders of radius $0.5$ kpc.
To measure the pattern speed at the time of the output $n$, we compare the principal axes between the outputs $n-1$ and $n+1$.

The evolution of $\Omega_\mathrm{p}$ is presented in Figure \ref{fig_omega_p}. Each curve starts shortly after the first
pericenter passage, when the bar is fully formed. Soon after formation all bars have a similar pattern speed of
$\sim 4$ Gyr$^{-1}$, however the detailed hierarchy corresponds to the reversed order of strength, as could have been expected from previous studies \citep{athanassoula03}. Later on we can
observe short-term variation, especially at the pericenter passages, when the bars are highly distorted and $A_2$ is
dominated by the elongation due to the proximity of the host. The tidal force also exerts a torque on the bar, as
discussed by \citet{lokas14}, which may affect the bar rotation in the vicinity of the pericenter.
We verified that the changes of the pattern speed are not driven by the mass loss. The evolution of both stellar and dark matter mass in the bar region is very similar for the three dwarfs on the same orbit (M0, M45 and M90), yet their $\Omega_\mathrm{p}$ varies differently.

Apart from short-timescale fluctuations, we notice also long-term trends.
In case of the in-plane prograde dwarfs on small-pericenter
orbits (S0 and M0), bars are significantly slowing down and finally halt after a few pericenters. Interestingly, the bar in
run M0 temporarily reverses its rotation speed around $7$ Gyr. On the other hand, the slowdown is more gradual in L0 and M90 runs.
In M45 the pattern speed appears to be on average constant. In all cases, the evolution of $\Omega_\mathrm{p}$ has trends similar to those of the
mean rotation velocity change shown in Figure~\ref{fig_rotation}.

The most obvious property of a bar is its length. \citet{athanassoula_misiriotis02} discussed various methods to
estimate it. We note there is no method to measure the bar size unambiguously and many of them require setting some
\emph{ad hoc} parameters. We decided to employ a method inspired by the aforementioned paper. We define the bar
semi-major axis length as the distance from the center where triaxiality $T$ obtained from our algorithm drops below
$90\%$ of its maximal value. We motivate this choice by the fact that $T$ profiles are quite flat in the
center and further outside drop steeply. Our selected threshold intends to reflect this drop.

\begin{figure}[t!]
\includegraphics[width=\columnwidth]{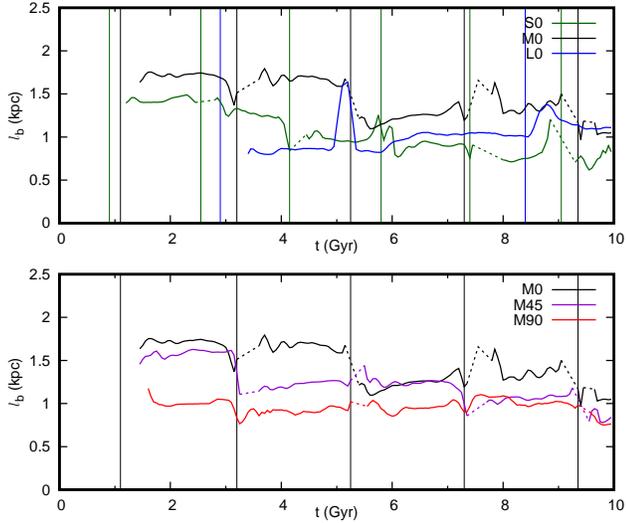}
\caption{The bar length $l_\mathrm{b}$, measured from the drop in triaxiality, as a function of time. To smooth the results, we applied a
moving average over three consecutive outputs. Dashed parts of the lines correspond to periods when the bar length
determination was not possible. The vertical lines indicate pericenter passages.}
\label{fig_bar_length}
\end{figure}

In Figure \ref{fig_bar_length} we depicted the evolution of the bar lengths. We note that at some outputs we were
unable to determine $l_\mathrm{b}$. The typical reason was that the dwarfs at pericenters, or shortly afterwards, were
so disturbed that our method did not work, mainly because $T$ grew all the way from the center to the outskirts. As
was the case for many other properties discussed so far, $l_\mathrm{b}$ is quite stable between pericenters, while it
usually changes after each pericenter passage. Hence, the bars do not undergo significant secular evolution.

The longest bar was formed in dwarf M0, whereas in runs with smaller (S0) and larger (L0) orbit size the bar was
initially shorter. Furthermore, there is a trend with disc inclination, such that for a larger one the bar is
shorter. However, the evolution of $l_\mathrm{b}$ is not very easy to predict. For example, sometimes the M45 bar
has the length equal to the one of M0, but at other instances equal to the one of M90.
We note that at the end of simulations (except for L0) the dwarfs should be regared as having elongated central parts, surrounded by spheroidal envelope, rather than as a disc with a bar component.
The sudden increase and then decrease of L0 bar length around $5$ Gyr is caused by the winding up of spiral arms and
corresponds to an elongated feature visible in Figure~\ref{fig_density_maps}.

\subsection{Dynamics of the bar rotation}

\begin{figure}
\includegraphics[width=\columnwidth]{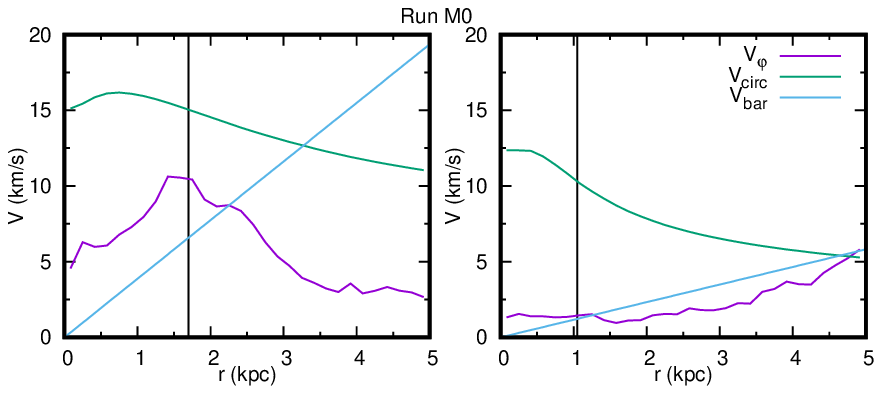} \\
\includegraphics[width=\columnwidth]{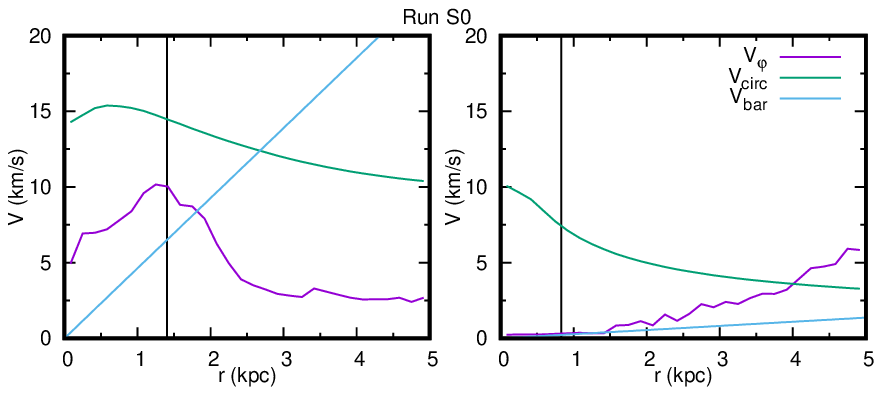} \\
\includegraphics[width=\columnwidth]{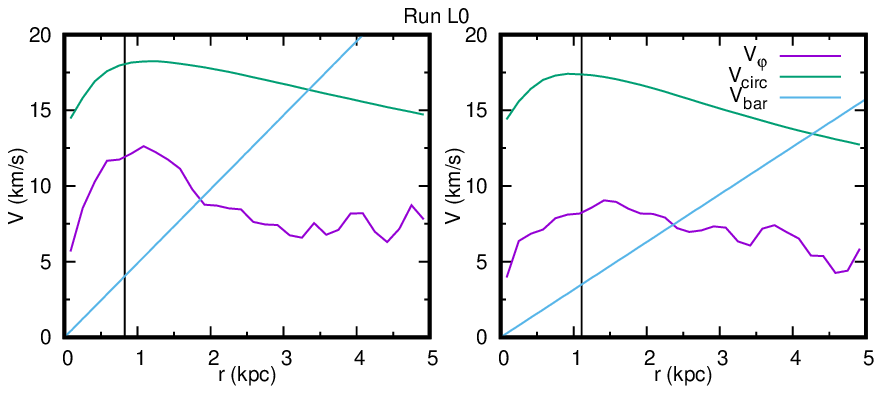} \\
\includegraphics[width=\columnwidth]{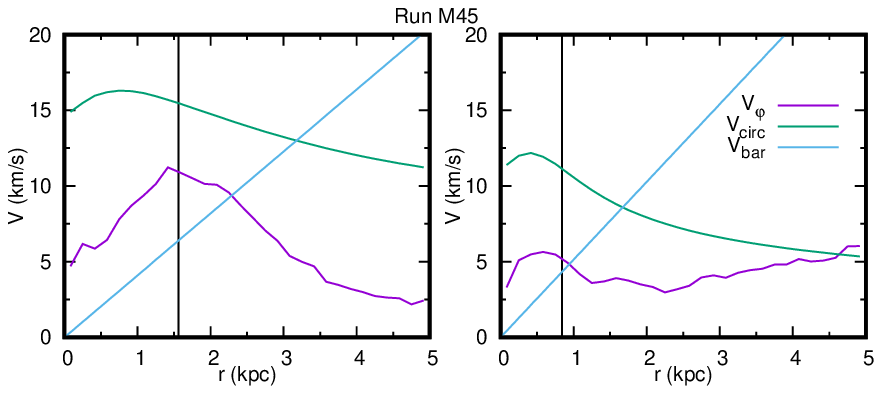} \\
\includegraphics[width=\columnwidth]{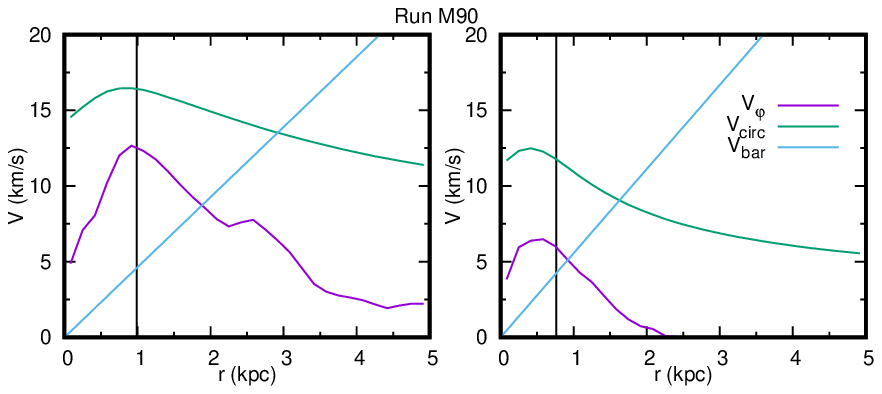}
\caption{The circular velocity $V_\mathrm{circ}$ (green), the true rotation velocity $V_\varphi$ (violet) and the bar
speed at a given distance $V_\mathrm{bar}=r\Omega_\mathrm{p}$ (blue). Vertical lines indicate the bar length $l_\mathrm{b}$.
Left column depicts the second apocenter and the right column the final state.}
\label{fig_cr}
\end{figure}

An important parameter for the bar dynamics is a ratio of the corotation radius to the bar length
$\mathcal{R}=R_\mathrm{CR}/l_\mathrm{b}$. The corotation radius $R_\mathrm{CR}$ is defined as
$\Omega_\mathrm{circ}(R_\mathrm{CR})=\Omega_\mathrm{p}$, where $\Omega_\mathrm{circ}(R)$ is the circular velocity of
the galaxy.

A careful analysis of possible orbits in bars by \citet{contopoulos80} showed that bars must follow the relation $R_\mathrm{CR} > l_\mathrm{b}$ (i.e. $\mathcal{R}>1$).
A similar conclusion was reached by \citet{athanassoula80}, who used forcings of different extents but found that bars were always shorter than the corotation radius.
This limit to the bar length can be illustrated by a following simplified reasoning.
Most of the stellar particles in the bar are on prograde orbits (i.e. they orbit around the dwarf center faster than the
bar rotates). However, the velocity of the particles at apocenters cannot be larger than the circular velocity.
The pattern speed of the bar is the same at every radius, hence the bar must end before circular angular speed drops
below it, meaning that we expect $R_\mathrm{CR} > l_\mathrm{b}$.

Bars with $\mathcal{R}<1.4$ are considered \emph{fast}, because their pattern speed is close to the maximal possible rotation speed at a given bar length.
Conversely, bars for which $\mathcal{R}>1.4$ are called \emph{slow}.
In general, $N$-body simulations indicate that bars formed in isolation tend to be fast and indeed have $1<\mathcal{R}<1.4$ \citep[see][]{athanassoula13}.
However, this may depend to some extent on the choice of initial parameters as slow bars have also been shown to form in isolation \citep{lokas16}.
Bars formed in interactions are rather slow \citep[e.g.][]{miwa_noguchi98}.

We analyzed in detail the rotation in the dwarfs. However, we present it in terms of the rotation velocity instead of
angular speed, because this way highlights more features. In Figure \ref{fig_cr} we plotted the circular velocity
$V_\mathrm{circ}$, the true rotation velocity of the stellar component $V_\varphi$ and the linear bar rotation velocity
$V_\mathrm{bar}=R\Omega_\mathrm{p}$. We computed the circular velocity as
\begin{equation}
	V_\mathrm{circ}(R)=\left[\frac{GM(r<R)}{R}\right]^{1/2},
\end{equation}
where $G$ is the gravitational constant, $r$ is the distance from the center and $M(r<R)$ is the total mass (both
baryonic and dark) closer to the dwarf's center than $R$. We calculated $V_\varphi$ by averaging particle velocities at
a given distance. The bar length is also indicated in each panel.

The much lower values of $V_\mathrm{circ}$ at the end of simulations are the effect of mass stripping, mainly of the dark
matter component. 
The growth of $V_\varphi$ in the outskirts, visible in some panels, is related to the transition to the tidal arms.
In some cases (e.g.\ at the end of run S0) $V_\varphi$ is larger than $V_\mathrm{circ}$.
This is not surprising, as particles located in that region are already stripped from the dwarf galaxy and their movement is governed by the potential of the host.
The sizes of the dwarfs can be estimated from Figure \ref{fig_density_maps} and at the end of the simulations are smaller than $2$ kpc (except L0, which is larger).
The particles already stripped from the dwarf can
move with a significant relative velocity \citep{lokas13}.

The maximum of the initial rotation velocity curve was around $19.4$ km s$^{-1}$, attained approximately $2$ kpc from
the center. As already discussed (see Figure~\ref{fig_rotation}), in all cases the dwarfs lost a significant fraction
of rotation, especially in the outer parts, but also in the centers. The loss of rotation is caused of course
by the tidal force and stripping. The shapes of $V_\varphi(R)$ are qualitatively similar for all our simulations, usually with a distinct
maximum. Interestingly, the measured bar lengths fall in the vicinity of these maxima. However, recall that measuring
$l_\mathrm{b}$ involves a free parameter, hence it may be only a coincidence.

The values of $\mathcal{R}$ can be obtained by finding the intersection of $V_\mathrm{bar}$ and $V_\mathrm{circ}$ (blue
and green) curves and comparing them to the bar length. We conclude that all our bars are slow, having values of
$\mathcal{R}$ from $2$ to $4$ and more. However, considering the true rotation velocity $V_\varphi$ (violet lines), we
see that the bars are actually rotating quite fast compared to the rotation velocity of the stars. By similar argument
as above, bars are limited by the \emph{true corotation}, i.e. the radius where the pattern speed is the same as the
actual angular velocity of the stars. 

Out of the in-plane prograde dwarfs, the slowest (in terms of $\mathcal{R}$) is the bar in run L0, which is caused by its short
length. However, one has to remember that if we consider the periods after the first pericenters, this bar
is actually \emph{the fastest} in terms of $\Omega_\mathrm{p}$. For the inclined dwarfs, $\mathcal{R}$ drops with
inclination, but here again M90 is the fastest in terms of the pattern speed.
Hence one has to be careful, since bars may be fast according to one definition and at the same time slow according to another \citep[see also][]{font17}.

At the end of the simulations, in the two in-plane prograde dwarfs with small pericenters (M0 and S0) the bars are very slow, as
they are barely rotating. The dwarfs with inclined discs retained their pattern speed (see Figure~\ref{fig_omega_p})
and at the same time they lost rotation and a lot of mass, hence they are actually faster than at the second apocenter.

\subsection{Angular momentum transfer}

\begin{figure}
\includegraphics[width=\columnwidth]{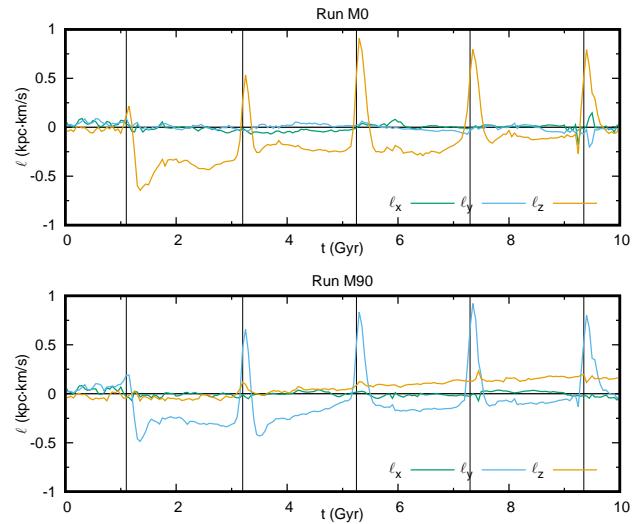}
\caption{The specific angular momentum $\ell$ of the central parts of the dark matter haloes in simulations M0 and M90 (see text) as a function of time. Here the reference frame is such that the initial
angular momentum of the disc is along the $z$ axis. In such a frame, the orbital angular momentum of dwarf M0 is also
along the $z$ axis, but in run M90 it is along the $y$ axis (i.e. the dwarf orbits in the $xz$ plane).}
\label{fig_angular_momentum}
\end{figure}

An important concept in the studies of barred galaxies is the angular momentum transfer. It is quite well established
that in the case of simulations of isolated galaxies the bars emit angular momentum, which is absorbed
by their dark matter haloes. This process enables bars to grow in the secular evolution phase. We intended to study this
process in the case of our simulations. Unfortunately, the stellar component loses a large fraction of its amount of angular
momentum via tidal interactions with the host galaxy directly to the tidal tails, and this leads to the decrease of
rotation in the whole disc.
Thus, any bar-related evolution of angular momentum was obscured by this process.

We were, however, able to track the behavior of the angular momentum of the dark matter halo. To study this effect, we
use a reference frame different from the one used previously. Now the $z$ axis is placed along the initial axis of
rotation of the disc and the axes are kept constant. In Figure~\ref{fig_angular_momentum} we show the time evolution of the specific angular
momentum $\ell$ of the central parts of dark matter haloes in two examples, M0 and M90. In the calculation we take into account all
particles closer than $2$ kpc from the centers of the dwarfs. Recall that in run M0 the dwarf was on the in-plane prograde
orbit, hence the $z$ axis is also perpendicular to the initial orbital plane. In run M90 the disc was perpendicular to
the orbital plane, so the $z$ axis lies \emph{in} this plane. The axis perpendicular to the plane is denoted as $y$.
The angular momentum evolution in runs S0 and L0 is qualitatively similar to run M0. In run M45 the disc is precessing
significantly, making the relevant plot incomprehensible.

The most prominent feature in both panels of Figure \ref{fig_angular_momentum} is the variation of angular momentum
component along the axis of the orbital motion ($z$ in run M0 and $y$ in run M90). After the first pericenter both halos
acquire negative (i.e. retrograde) angular momentum. This is a direct result of the preferential stripping of particles
on prograde orbits, in contrast to the retrograde ones \citep[and references therein]{henon70, gajda_lokas16}. The spikes appearing just
after the pericenter passages are due to the disturbance by the tidal force. In run M90 we can observe small, but
steady increase of the $\ell_z$ component, resulting from the angular momentum transfer. In the case of run M0, such an
increase in $\ell_z$ is buried under larger changes induced by tides.
Most of the angular momentum lost by the stellar component is not transferred to the dark matter halo of the dwarf galaxy.

\section{Discussion}
\label{sec_discussion}

\subsection{Buckling instability and thickening}

Buckling instability is a common phase of bar evolution. In the case of bars formed in isolation, it usually takes place
soon after the bar formation. The instability leads to a change in the velocity distribution of the galaxy. Obviously, the
vertical velocity dispersion $\sigma_z$ grows, but also the in-plane dispersion, e.g.\ $\sigma_R$, decreases
\citep[e.g.][]{martinez_valpuesta_shlosman04}. It has been argued \citep[see][and references therein]{bt08} that a critical parameter for the
instability is the ratio of $\sigma_z/\sigma_R$.

To check whether the dwarfs underwent buckling instability we investigated edge-on surface density maps. In runs M45,
M90 and L0 we did not detect any significant vertical asymmetry at any time. On the other hand, dwarf M0 underwent
buckling and in run S0 we also detected some weak traces of possible instability. Traces of undergoing buckling in
run M0 can be noticed in the $c/a$ measurement in Figure~\ref{fig_shape}, where it grows steadily between $3$ and $5$
Gyr. Moreover, $\sigma_\theta$ grows at the same time.

The occurrence of buckling in the simulations may be related to the strength of the individual bars. Dwarf M0 had the
strongest bar, with the highest level of velocity dispersion $\sigma_r$. The bar in run S0 was slightly weaker and in
run L0 was significantly weaker, at least after the first pericenter passage. This relation between the thickening amplitude and
bar strength is supported by the correlation in \citet{athanassoula08} and the simulations of \citet{martinez_valpuesta17}, among which a discernible boxy/peanut
bulge is present only in the case of the most disturbed galaxy with the strongest bar. In the case of the inclined
dwarfs, buckling might have been inhibited by the vertical disturbance of the disc during the pericenter passages.

One more reason why our bars did not experience strong buckling, might have been repeated pericenter passages. The dwarf
galaxies were tidally shocked at each of them and the resulting disturbance might have reduced the buckling amplitude.
Moreover, we cannot rule out that in cases where we did not notice the instability it actually occurred but was
too weak or too fast to be recognized.
None of our dwarfs formed a~boxy/peanut bulge, which is usually associated with the buckling instability.
Interestingly, \citet{erwin_debattista13} found that about $13\%$ galaxies in their sample do not show signs of boxy/peanut structures.
Moreover, the fraction of galaxies hosting a~boxy/peanut bulge is lower for the less massive systems \citep{erwin_debattista17}.

\subsection{Tidally induced spiral arms}

We would like to point out the development of spiral arms in our simulations, provided the tidal force is not too
strong. In run L0 we observe two sets of two-fold tightly wound spiral arms, both of which can be seen as distinct maxima in
the relevant panel of Figure~\ref{fig_a2_map}. The first one is located at $\approx 1.3$ kpc and exists for a period of
$4$-$6$ Gyr. The second is located at $\approx 2$ kpc and persists between $5$-$8$ Gyr. Tightly wound spiral arms are also present in our run M90 ($r\approx 1.2$ kpc, $t\approx2.5$-$3$ Gyr) and are probably responsible for the weakening of
the bar right after the second pericenter.

The inner set of arms can be linked to the peculiar features of L0 run around $5$ Gyr, such as peaks in the pattern
speed (Figure~\ref{fig_omega_p}) and the bar length (Figure~\ref{fig_bar_length}). Judging from the surface density
maps, at this time the spiral arms seem to wind up and weaken the bar. They detach from the bar ends and finally turn
into a ring around the bar, which can be seen in Figure~\ref{fig_density_maps}. A similar structure can be also seen in
some large spirals. 

The reader can refer to \citet{semczuk17} for a~detailed study of tidally induced spiral arms, albeit in the context of
normal-size galaxies orbiting in a galaxy cluster.

\subsection{Angular momentum and corotation}

Angular momentum and its transfer play an important role in the development and evolution of bars, as concluded from
works concerning bars formed through instability. However, in the case of tidally induced bars the angular momentum
budget is different. The galaxies encounter perturbers one or more times and lose large amounts of angular
momentum. Moreover, their outer parts are heavily disturbed and stripped.

Bars in isolated galaxies grow by emission of the angular momentum from inside the corotation radius, which is absorbed by the halo and
the outer part of the disc \citep{athanassoula03}. In our simulations $R_\mathrm{CR}$ is usually located in the outer part of the disk. In
particular, after the first pericenter $R_\mathrm{CR} \sim 3$ kpc, whereas initially $90\%$ of the disc mass is inside
$1.7$ kpc and the radius of $3$ kpc encompasses $99\%$ of the disc mass. Therefore, there was insufficient amount of
disc mass to absorb the angular momentum. In addition, our disc is initially hot and stable against bar formation.
\citet{athanassoula03} found that the hotter the disc the slower the deceleration of the bar.

In isolated galaxies, resonances of the dark matter halo can absorb angular momentum \citep{athanassoula03}. However, haloes in our simulations are heavily
stripped at all radii and heavily perturbed. On the fiducial orbit, the size of the halo drops from $6$ kpc after the first pericenter to $3$
kpc at the end, as can be estimated from the break in the dark matter density profile (see Figure 1 in
\citealt{gajda_lokas16}). Moreover, the halos lose mass also from the inner parts. Almost $60\%$ of the initial mass
inside $3$ kpc is lost at the first pericenter and more than $90\%$ before the end of the simulation. In addition, we
can imagine that the velocity distribution is perturbed, which may hamper angular momentum transfer
\citep{athanassoula03}.

The bars formed through rapid instability are usually fast, having $\mathcal{R}<1.4$ (i.e. $l_\mathrm{bar}>0.7
R_\mathrm{CR}$), which means that their length almost fills the corotation radius. Tidally induced bars however, are much
slower and/or shorter. Figure~\ref{fig_cr} makes it easy to understand, at least in the case of dwarf
galaxies and strong tidal forces. During the interaction, the rotation velocity $V_\varphi(R)$ drops significantly,
e.g. by a factor of 2. Hence, the bar cannot rotate as fast as in the isolated case. If we consider the \emph{true
corotation} (i.e. radius at which $V_\varphi/R=\Omega_\mathrm{p}$), then the bars appear to rotate almost as fast as possible,
given their environment.

We note that some bars formed in isolation may be slow (\citealt{villa_vargas09, lokas16}; gas-poor examples of \citealt{athanassoula14}),
possibly due to the lack of gas (including during the bar formation period), to a dominant dark matter halo, or when the corotation is outside the disc. We would like to note that
very slow pattern speeds, with an $\mathcal{R}$ ratio of the order of 3, have been actually observed in at least one case of a real galaxy \citep{elmegreen98}, while $\mathcal{R}$ ratios of the order of 2 are fairly common \citep[see][and references therein]{font17}.

Interestingly, the bar lengths in our simulations fall close to the maximum of the rotation curve and we cannot offer
any explanation for this, it may be only an accident.
However, it would be interesting to check how this relation looks in bars formed via instability and in less perturbed tidally induced bars.
It might be possibly related to the results of \citet{lynden_bell79}.

\subsection{Comparison with other works}

Now we would like to put our work in perspective of earlier findings. First we are going to compare our bars to bars
formed in isolation and then to other works on tidally induced bars.

We would like to stress that the evolution of bars in our simulation is different than in the case of bars formed
through instability. In the latter, bars usually start growing soon after the beginning of the simulation, then they
experience buckling instability and finally they continue growing in a secular fashion \citep[see][for a review]{athanassoula13}. Here, the initial model of the
dwarf galaxy is constructed in such a way that it is stable against spontaneous bar formation, at least over the timescales considered here (i.e. $10$ Gyr).
The bars develop during
the first encounter with the host galaxy. Later, however, they seem not to grow in a secular way, but rather
keep their parameters constant between pericenter passages. Usually, the repeated encounters with the host galaxy
weaken the bars. The only exception is the L0 dwarf on the widest orbit, for which the tidal force is relatively weak.

The profiles of the bar amplitude $|A_2|$ are also different with respect to bars formed in isolation. In the latter case
there is usually only a wide peak in the center, followed by a drop to near
zero in the outskirts \citep[e.g.][]{athanassoula02,
athanassoula03}. The tidally induced bars may exhibit many peaks in $|A_2|$ profiles. Some of them correspond to
tidally induced spiral arms, other result form shells of matter. Finally, $|A_2|$ grows in the outskirts, indicating
the transition from the main body of the galaxy to its tidal tails.

\citet{lokas16} analyzed a suite of simulations following a Milky Way-like galaxy orbiting in a
galaxy cluster. Their model of the progenitor galaxy was unstable to spontaneous bar formation, however,
it needed a long time to start developing the bar on its own. In their simulations the apparent influence of the tidal
force was much weaker than in our simulations, as can be judged from the $|A_2|$ enhancement at pericenter passages or
the tidal radius estimates. Consequently, the initially formed bars are weaker than in our simulations. However, later
on they grow continually, both during subsequent pericenter passages and in the periods between them. The
\citet{lokas16} bars do experience buckling instability, leading to the formation of boxy/peanut shapes in edge-on
views, contrary to our models.

\citet{janz12} found a bar fraction of $18\%$ among early-type dwarf galaxies in the Virgo cluster.
This may seem at odds with our simulations, in which all of the dwarfs developed bars.
However, there are various reasons why we do not observe $100\%$ bar fraction.
Firstly, presence of gas in real galaxies hampers bar formation \citep{athanassoula_et_al13}. Secondly, \citet{smith15} showed that in the case of galaxy clusters, dwarf galaxies need to plunge deep into the cluster core to be influenced significantly by the tidal forces.
Thirdly, bars in our simulations are weakened or even destroyed during the evolution.
As can be inferred from Figure \ref{fig_density_maps}, some of the dwarfs may not be classified as barred at final stages of their evolution, especially in case of unfavourable orientation.
Last but not least, we should note that all our simulations have the same mass model. Thus the fact that they all form bars cannot be compared to the observed fraction of bars in dwarfs galaxies, whose mass models span a considerable parameter space.

\section{Summary}
\label{sec_summary}

The purpose of this work was to study the formation and evolution of tidally induced bars in dwarf galaxies orbiting
a Milky Way-like host. We constructed an $N$-body model of a dwarf galaxy and verified that in isolation it is stable
against bar formation over the time ranges of relevance here. In order to induce bar formation, we put the dwarf galaxy on an elongated orbit around an
$N$-body model of a galaxy resembling the Milky Way. We focus on two parameters governing the evolution: the size of the
dwarf's orbit and its disc inclination with respect to the orbital plane. We conducted five simulations. Three of them
had an in-plane prograde orientation of the disk but varying orbit sizes. For the medium-sized orbit we performed two additional
simulations with the dwarf's disc inclined by $45\degr$ and $90\degr$ with respect to the orbital plane.

In all cases, bars form during the first pericenter passage, however their properties and the subsequent evolution
vary. The overall strongest bar was formed for the intermediate orbit and in-plane prograde disc orientation. On the tighter
orbit the tidal force is stronger and the disc is disturbed excessively, inhibiting the formation of a strong bar. On
the wider orbit the tidal forces are too weak, hence initially only a weak bar develops. In comparison to a in-plane prograde
orbit, for the higher inclinations the bars are progressively weaker. In the retrograde case, \citet{lokas15} found
that no bar forms at all.
The dependence on the disc orientation stems from the progressively weaker tidal force in the equatorial plane of the dwarf and the unfavorable orientation of the Coriolis force.

The bars formed in our simulations were slow ($\mathcal{R}\sim 2-3$), in contrast to the bars developed in isolation,
which are typically classified as fast ($\mathcal{R}<1.4$). We found that this is caused by the loss of a huge fraction
of the dwarf galaxy angular momentum during the pericenter passage, as initially proposed by \citet{miwa_noguchi98}.
Consequently, the stellar component of the dwarf was rotating much slower than the maximum allowed by the circular
velocity. Hence, the bars were rotating significantly slower than the maxima deduced from the comparison of their lengths to their circular velocity curves.

The life of bars developed in isolation usually includes phases of buckling and secular evolution. Here,
even if a bar underwent buckling, it was only very weak.
One possible reason is that the bars were weaker than in the isolated cases \citep{athanassoula08}.
We also did not observe noticeable secular evolution, which usually involves the bar growth and
slowdown. The possible explanation is the combination of repeated tidal shocks and stripping, which hindered the angular
momentum transfer.

The bars were stable between pericenter passages and evolved mainly during them. Except for the case of the widest
orbit, during subsequent encounters with the host the bars were weakened and shortened. In addition, the in-plane prograde
dwarfs on the tightest and the intermediate orbit quickly lost almost all of their rotation. The final states of those
dwarfs consisted of an elongated central part, with some rotation possible, enshrouded in a more spheroidal envelope.
Such an evolution was envisaged by the tidal stirring scenario for the transformation of a disc into a spheroid \citep{mayer01, kazantzidis11}. Obviously, it is not the buckling which leads to this thickening, but rather tidal shocking.

The main concern regarding the applicability of this work to the dwarf galaxies e.g.\ in the Local Group is whether the
initial conditions of the simulations correspond to real systems prior to accretion into the neighborhood of a larger
galaxy. Dwarf galaxies observed far away from other galaxies often posses significant amount of rotating interstellar
gas. However, it remains an open question if their stellar components have structures of rather thin discs rotating in
an ordered fashion, as we assumed here.
Moreover, in reality the satellites are accreted from the cosmic web, whereas here we just placed the dwarfs at apocenters of their orbit and gave them appropriate speed to reach the desired pericenter.
Furthermore,
we neglected entirely the gas component and performed fully
collisionless simulations. We know that the collisional component impacts immensely the formation and evolution of
bars in isolation \citep{athanassoula92b, villa_vargas10, athanassoula_et_al13, athanassoula14} and that in fact isolated dwarf galaxies
contain a large gas fraction \citep{papastergis12, sales15}. Although the gas is probably efficiently ram-pressure
stripped by the hot halo of the host, it may remain in the dwarf long enough to significantly affect the evolution. We
intend to study the influence of the interstellar medium on the formation of tidally induced bars in a~future work.

\acknowledgments
This work was partially supported by the Polish National Science Centre under grant 2013/10/A/ST9/00023. EA is partially supported by the CNES (Centre National d'\'{E}tudes Spatiales, France).

\appendix

\section{Calculation of the pattern speed}
\label{appendix}
The purpose of this appendix is to solve the following problem.
At an initial time, normalised vectors $\bm{x}_\mathrm{i}$, $\bm{y}_\mathrm{i}$ and $\bm{z}_\mathrm{i}$ point along bar principal axes, respectively, major, intermediate and minor.
Similarly, vectors $\bm{x}_\mathrm{f}$, $\bm{y}_\mathrm{f}$ and $\bm{z}_\mathrm{f}$ correspond to a later state of this galaxy, which we will hereafter refer to as the final state of the galaxy.
We would like to give a proper definition of the angle $\omega$ by which the bar rotated around the minor axis and to calculate its value in terms of $\bm{x}_\mathrm{i}$, $\bm{y}_\mathrm{i}$, $\bm{z}_\mathrm{i}$ and $\bm{x}_\mathrm{f}$, $\bm{y}_\mathrm{f}$, $\bm{z}_\mathrm{f}$.

We note that all the vectors we discuss in this appendix are normalised to unity. For purposes of matrix operations, vectors should be understood as columns consisting of three numbers.

Without loss of generality, we can choose a reference frame whose axes lie along principal axes of the bar at the initial time, i.e.\ $\bm{x}_\mathrm{i}'=\bm{\hat x}$ and so forth.
In such a~frame, the final principal axes can be expressed as
\begin{equation}
\label{eq_ref}
\bm{x}_\mathrm{f}'= 
\left(
	\begin{array}{c}
	\bm{x}_\mathrm{i}^\mathsf{T} \\
	\bm{y}_\mathrm{i}^\mathsf{T} \\
	\bm{z}_\mathrm{i}^\mathsf{T}
	\end{array}
\right) \bm{x}_\mathrm{f},
\end{equation}
where $^\mathsf{T}$ stands for transposition, so the first object on the right-hand side should be understood as a matrix, whose rows are given by components of $\bm{x}_\mathrm{i}$, $\bm{y}_\mathrm{i}$ and $\bm{z}_\mathrm{i}$.
Similar expressions can be written for $\bm{y}_\mathrm{f}'$ and $\bm{z}_\mathrm{f}'$.
The transformation of $\{\bm{x}_\mathrm{i}', \bm{y}_\mathrm{i}', \bm{z}_\mathrm{i}'\}=\{ \bm{\hat x}, \bm{\hat y}, \bm{\hat z} \}$ set into $\{\bm{x}_\mathrm{f}', \bm{y}_\mathrm{f}', \bm{z}_\mathrm{f}'\}$ can be described in terms of a matrix $\mathbf{T}$ such that
\begin{subequations}
\label{eq_def}
\begin{eqnarray}
\mathbf{T}\bm{x}_\mathrm{i}'&=\mathbf{T}\bm{\hat{x}}=\bm{x}_\mathrm{f}', \\
\mathbf{T}\bm{y}_\mathrm{i}'&=\mathbf{T}\bm{\hat{y}}=\bm{y}_\mathrm{f}', \\
\mathbf{T}\bm{z}_\mathrm{i}'&=\mathbf{T}\bm{\hat{z}}=\bm{z}_\mathrm{f}',
\end{eqnarray}
\end{subequations}
We will seek a relation between $\mathbf{T}$ and $\omega$.

Let us introduce the following notation: $\mathbf{R}(\bm{k}, \phi)$ will be a matrix, which is an operator rotating vectors about unit vector $\bm{k}$ by angle $\phi$ in counter clockwise direction.
Elements of such a matrix can be found using the Rodrigues' rotation formula:
\begin{equation}
\mathbf{R}(\bm{k}, \phi)=\mathbf{I}+\mathbf{K}\sin\phi+\mathbf{K}^2(1-\cos\phi),
\end{equation}
where $\mathbf{I}$ is the identity matrix and $\mathbf{K}$ is a following matrix 
\begin{equation}
\mathbf{K}= \left(
\begin{array}{ccc}
  0   &  -\bm{k}_z  &   \bm{k}_y \\
 \bm{k}_z  &    0   &  -\bm{k}_x \\
-\bm{k}_y  &   \bm{k}_x  &    0 
\end{array}
\right).
\end{equation}
Matrix $\mathbf{R}(\bm{k},\phi)$ can be viewed as an active transformation, as we described above, but as well as a passive one.
Namely, it corresponds to a change of reference frame, such that the frame is rotated about $\bm{k}$ by an angle $-\phi$.
From construction, $\mathbf{R}$ is an orthogonal matrix and its inverse equals to $\mathbf{R}(\bm{k},\phi)^{-1}=\mathbf{R}(\bm{k},\phi)^\mathsf{T}=\mathbf{R}(\bm{k},-\phi)$.

We need to make an assumption regarding the trajectory of the bar minor axis. 
This is required because otherwise we would not be able to determine a ``zero-point'' of the bar rotation angle.
In particular, if the path of $\bm{z}$ was unrestricted and the bar was not rotating, it could have any orientation at the final time.
We assume the simplest possibility, namely that the bar minor axis follows the shortest path between $\bm{\hat{z}}$ and $\bm{z}_\mathrm{f}'$.
It can be regarded as a good approximation, if the outputs from the simulations are dense enough. Hence, the transformation for $\bm{z}$ is following:
\begin{equation}
\bm{z}_\mathrm{f}'=\mathbf{R}(\bm{n},\theta)\bm{\hat{z}},
\end{equation}
where $\bm{n}=\frac{\bm{\hat{z}}\times\bm{z}_\mathrm{f}'}{|\bm{\hat{z}}\times\bm{z}_\mathrm{f}'|}$ and $\theta=\arccos(\bm{\hat{z}}\cdot\bm{z}_\mathrm{f}')=\arccos(\bm{z}_\mathrm{i}\cdot\bm{z}_\mathrm{f})$.

One natural possibility for the full transformation matrix is $\mathbf{T}=\mathbf{R}(\bm{n},\theta)\mathbf{R}(\bm{\hat{z}},\omega)$, i.e.\ first we rotate about $\bm{\hat{z}}$ by $\omega$ and then about $\bm{n}$ by $\theta$.
However, there is another possibility, namely $\mathbf{R}(\bm{z}_\mathrm{f}',\omega)\mathbf{R}(\bm{n},\theta)$, i.e.\ first we rotate about $\bm{n}$ by $\theta$ and then about $\bm{z}_\mathrm{f}'$ by $\omega$.
It turns out they are equivalent, as can be seen considering an operation $\mathbf{A}=\mathbf{R}(\bm{n},\theta)\mathbf{R}(\bm{\hat{z}},\omega)\mathbf{R}(\bm{n},-\theta)$.
It can be regarded as composition of (i) rotation of the reference frame by $\theta$ about $\bm{n}$, which sets the $z$-axis along $\bm{z}_\mathrm{f}'$ (ii) rotation around new $z$-axis and (iii) rotation of the reference frame back to the original position.
Hence, it is obvious that in fact $\mathbf{A}=\mathbf{R}(\bm{z}_\mathrm{f}',\omega)$, from which follows that $\mathbf{R}(\bm{n},\theta)\mathbf{R}(\bm{\hat{z}},\omega)=\mathbf{R}(\bm{z}_\mathrm{f}',\omega)\mathbf{R}(\bm{n},\theta)$.
It can be also checked through tedious, but straightforward calculation of the elements of the matrices.

Using the above equality, we can think about the whole transformation as composed of infinitesimal rotations about $\bm{n}$, followed by rotation around the instantaneous $z$-axis, repeated until the final state is reached.
Hence, $\mathbf{T}$ can be thought of as
\begin{equation}
\mathbf{T}=\prod_{j=0}^{N-1} \mathbf{R}(\bm{n},\theta/N)\mathbf{R}( \bm{z}_j,\omega/N),
\end{equation}
where $\bm{z}_j=\mathbf{R}(\bm{n}, \theta/N)^j\bm{\hat{z}}=\mathbf{R}(\bm{n}, \theta j/N)\bm{\hat{z}}$.

The general structure of the matrix $\mathbf{T}=\mathbf{R}(\bm{n},\theta)\mathbf{R}(\bm{\hat{z}},\omega)$ is complicated, however combination of its elements takes a simple form
\begin{equation}
\mathbf{T}_{21}-\mathbf{T}_{12}=(1+\cos\theta)\sin\omega.
\end{equation}
Moreover, $\mathbf{T}_{33}=\cos\theta$.
Hence, the rotation angle $\omega$ can be obtained from the numerical representation of $\mathbf{T}$ using
\begin{equation}
\omega=\arcsin\left( \frac{\mathbf{T}_{21}-\mathbf{T}_{12}}{1+\mathbf{T}_{33}} \right).
\end{equation}

Let us now turn to the determination of $\mathbf{T}$ from the initial and final orientation of the bar principal axes. Combining the definitions in \eqref{eq_def} we get
\begin{equation}
(
	\begin{array}{ccc}
	\bm{x}_\mathrm{f}' & \bm{y}_\mathrm{f}' & \bm{z}_\mathrm{f}'
	\end{array}
)
=\mathbf{T}
(
	\begin{array}{ccc}
	\bm{\hat{x}} & \bm{\hat{y}} & \bm{\hat{z}}
	\end{array}
)=\mathbf{T}
\end{equation}
Hence, the columns of $\mathbf{T}$ are given by components of $\bm{x}_\mathrm{f}'$, $\bm{y}_\mathrm{f}'$ and $\bm{z}_\mathrm{f}'$.
We can further substitute the expression for $\bm{x}_\mathrm{f}'$ from \eqref{eq_ref} and similarly for $\bm{y}_\mathrm{f}'$ and $\bm{z}_\mathrm{f}'$. Finally,
\begin{equation}
\mathbf{T}=
\left(
	\begin{array}{c}
	\bm{x}_\mathrm{i}^\mathsf{T} \\
	\bm{y}_\mathrm{i}^\mathsf{T} \\
	\bm{z}_\mathrm{i}^\mathsf{T}
	\end{array}
\right)
(
	\begin{array}{ccc}
	\bm{x}_\mathrm{f} & \bm{y}_\mathrm{f} & \bm{z}_\mathrm{f}
	\end{array}
).
\end{equation}
From the above form one can easily read the appropriate components of $\mathbf{T}$, obtaining
\begin{equation}
\label{eq_final1}
\omega = \arcsin\left( \frac{\bm{y}_\mathrm{i}\cdot \bm{x}_\mathrm{f} - \bm{x}_\mathrm{i}\cdot \bm{y}_\mathrm{f}}{1+\bm{z}_\mathrm{i}\cdot \bm{z}_\mathrm{f} } \right).
\end{equation}
The orientation of the intermediate axis can be expressed using $\bm{y}=\bm{z}\times\bm{x}$. Thus, the above result can be rewritten as
\begin{equation}
\label{eq_final2}
\omega = \arcsin\left( \frac{
(\bm{z}_\mathrm{i} + \bm{z}_\mathrm{f}) \cdot ( \bm{x}_\mathrm{i} \times \bm{x}_\mathrm{f}) }{1+\bm{z}_\mathrm{i}\cdot \bm{z}_\mathrm{f} }
\right).
\end{equation}
\vspace{0.1pt}

\noindent
If the disc is not precessing (i.e. $\bm{z}_\mathrm{f}=\bm{z}_\mathrm{i}=\bm{z}$), it simplifies to
\begin{equation}
\omega=\arcsin[ \bm{z}\cdot ( \bm{x}_\mathrm{i} \times \bm{x}_\mathrm{f}) ],
\end{equation}

which can be easily understood and serves as a simple check for our considerations. We see that $\omega\in[-\pi/2,\pi/2]$ is an oriented angle, which is positive if the bar rotates counter clockwise around its minor axis and negative otherwise.

\end{document}